\documentclass[fleqn,usenatbib]{mnras}

\usepackage{newtxtext,newtxmath}
\usepackage{verbatim}
\usepackage{pdflscape}
\usepackage{comment}
\usepackage[T1]{fontenc}
\usepackage{lipsum}

\DeclareRobustCommand{\VAN}[3]{#2}
\let\VANthebibliography\thebibliography
\def\thebibliography{\DeclareRobustCommand{\VAN}[3]{##3}\VANthebibliography}

\usepackage{graphicx}	% Including figure files
\usepackage{amsmath}	% Advanced maths commands
\usepackage{multirow}
\usepackage[dvipsnames]{xcolor}
\usepackage[pscoord]{eso-pic}
\usepackage[normalem]{ulem}
\usepackage{makecell}
\usepackage{xspace}
\usepackage{dcolumn}

\newcommand{\CosmoSpec}{{\tt CosmoSpec}\xspace}
\newcommand{\CosmoTherm}{{\tt CosmoTherm}\xspace}

\newcommand{\LCDM}{$\Lambda$CDM\xspace}
\newcommand{\Planck}{{\it Planck}\xspace}

\newcommand{\tquote}[1]{``#1''}

\newcommand{\baromb}{\bar{\omega}_{\rm b}}
\newcommand{\omb}{\omega_{\rm b}}

\def\aap{A\&A}
\def\apj{ApJ}
\def\apjs{ApJS}
\def\apjl{ApJL}
\def\mnras{MNRAS}

\def\prd{Phys. Rev. D}         % Physical Review D
\def\apss{ApSS}

\title[CRR emulation and variance]{{\tt CRRfast}: An emulator for the Cosmological Recombination Radiation with effects from inhomogeneous recombination}

\author[Lucca et al. (2023)]{
	Matteo Lucca$^{1}$,
	Jens Chluba$^{2}$
	and
	Aditya Rotti$^{2}$
	\\\\
	$^{1}$Service de Physique Th\'eorique, Universit\'e Libre de Bruxelles, C.P. 225, B-1050 Brussels, Belgium\\
	$^{2}$Jodrell Bank Centre for Astrophysics, University of Manchester, Manchester M13 9PL, UK
}

\date{\vspace{-3mm}Accepted XXX. Received YYY; in original form ZZZ}

\pubyear{2023}

\begin{document}

\label{firstpage}
\pagerange{\pageref{firstpage}--\pageref{lastpage}}
\maketitle
	
\begin{abstract}
The Cosmological Recombination Radiation (CRR) is one of the guaranteed $\Lambda$CDM Spectral Distortion (SD) signals. Even if very small in amplitude, it provides a direct probe of the three recombination eras, opening the path for testing one of the key pillars in our cosmological interpretation of the measured CMB anisotropies. Here we develop 
a new emulator, {\tt CRRfast}, to quickly and accurately represent the CRR for a wide range of cosmologies, using the state-of-the-art \texttt{CosmoSpec} code as a reference. {\tt CRRfast} has been made publicly available both as stand-alone code and as part of \texttt{CLASS}, thereby {\it completing} the set of \LCDM sources of SDs that can be modeled with \texttt{CLASS}. With this newly-developed pipeline we investigate the full constraining power of SDs within \LCDM and highlight possible future applications to experimental design optimization. Furthermore, we show that the inhomogeneous evolution of the recombination process imprints second-order contributions to the CRR spectrum, leading to a broadening and shifting of the CRR features. These second-order terms are naturally captured by the emulator and allow us to evaluate the $\Lambda$CDM 
contributions to the average CRR
as well as to illustrate the effect of perturbed recombination due to Primordial Magnetic Fields (PMFs). As it turns out, while the \LCDM variance effects can be neglected, they could be significantly enhanced in the beyond-\LCDM models. 
In particular in the case of PMFs we demonstrate that through these non-linear terms the 
parameter space relevant to the Hubble tension could be tested with future CMB spectrometers. 
\end{abstract}

% Select between one and six entries from the list of approved keywords.
% Don't make up new ones.

%%%%%%%%%%%%%%%%%%%%%%%%%%%%%%%%%%%%%%%%%%%%%%%%%%

%%%%%%%%%%%%%%%%% BODY OF PAPER %%%%%%%%%%%%%%%%%%

\section{Introduction}
%------------------------------------

Well on our way into the era of precision cosmology, many of the assumptions underlying the standard cosmological model, \LCDM, have been put under the magnifying glass. One particularly notable example is the recombination process, which represents one of the key theoretical ingredients of \LCDM that might also help explain (at least in part) the origin of the Hubble tension \citep[see e.g.,][for recent reviews]{DiValentino2021Realm, Schoneberg2021Olympics}. Concretely, the presence of both Primordial Magnetic Fields \citep[PMFs --][]{Jedamzik2020Reducing, Galli:2021mxk} and Varying Fundamental Constants (VFC), such as the electron mass $m_{\rm e}$ and the fine-structure constant $\alpha$ \citep{Hart2019Updated, Hart:2021kad, Sekiguchi2021, Lee:2022gzh}, directly modify the recombination history in a way that is favoured over \LCDM for specific combinations of data sets, resulting in an overall reduction of the Hubble tension. Furthermore, models modifying the expansion history of the universe around the time of recombination, such as Early Dark Energy \citep[EDE -- see e.g.,][for recent reviews]{Hill2020Early, Poulin:2023lkg}, indirectly affect the recombination dynamics, being dictated by a balance between expansion and atomic interaction rates. As such, it cannot be excluded that recombination proceeds differently than what is expected within the \LCDM model.

One of the main reasons why there is such a (relatively) large freedom in the evolution of the recombination process is that the shape of the Cosmic Microwave Background (CMB) anisotropy power spectra (the most precise probe of the early universe available to date) is only sensitive to its integrated effect, most notably via the visibility function and the definition of the angular distance to the last-scattering surface. This is an intrinsic limitation of the CMB anisotropies as observable of the early universe and can only be fully overcome with the help of complementary information. However, there is one way to directly test the recombination history and that is by observing the Cosmological Recombination Radiation \citep[CRR --][]{Peebles1968Recombination, Dubrovich1975Hydrogen, Rubino2006, Rubino2008Lines, Chluba2006FF, Chluba2008There, Chluba2009Pre, Sunyaev2009Signals, AliHaimoud2013RecSpec, Chluba2016Cosmospec}. 

In brief, the CRR is a Spectral Distortion (SD) of the CMB energy spectrum that is sourced by the emission/absorption of photons during the recombination of helium and hydrogen. As a consequence, its shape inherits unique spectral features that have contributions from all stages of the recombination process. Therefore, its precise measurement would allow to directly test the time evolution as well as the exact characteristics of recombination \citep{Sunyaev2009Signals}, thereby opening a way to probe new physics \citep[see e.g.,][]{Chluba2009Pre,Chluba2010Could} and modifications of the expansion rate \citep{Hart2020Sensitivity}. Although the CRR has so far eluded observations \citep[as have primordial CMB SDs in more general, see e.g.,][]{Chluba2019Voyage}, up-coming experiments such as the ground-based Array of Precision Spectrometers for the Epoch of Recombination (APSERa) \citep{SathyanarayanaRao:2015vgh, SathyanarayanaRao:2016ope, 9726484} and the more advanced setups recently proposed as part of the Voyage 2050 initiative \citep[][referred to henceforth only as Voyage 2050 mission]{Vince2015Detecting, Chluba2019Voyage} could potentially detect the CRR signal with high significance.

Measurements of the CRR could therefore significantly deepen our understanding of the recombination process and shed light on the aforementioned variations from the standard picture. For instance, the role that the future observation of the CRR with a Voyage 2050 mission could play in the context of EDE and VFC has been recently discussed in \cite{Hart:2022agu}. The work carried out in this manuscript is meant to extend these previous analyses and highlight even more strongly the versatility of the CRR as a valuable probe of the recombination era.
In particular, here we develop a new numerical tool to evaluate the CRR spectrum within the \LCDM model and some of its minimal extensions. This emulator, named \texttt{CRRfast}, is based on the exact calculations of \texttt{CosmoSpec} \citep{Chluba2016Cosmospec}, which in turn is based on the recombination code \texttt{CosmoRec} \citep{Chluba2011Towards}. The idea behind the emulator is to \textit{i)} Taylor-expand the CRR spectrum around a fiducial for all relevant cosmological parameters, \textit{ii)} tabulate and save the resulting Taylor coefficients and \textit{iii)} use the latter to calculate the CRR for any set of parameter values. 
In this way, \texttt{CRRfast} is significantly faster than \texttt{CosmoSpec} (by a factor of more than $500$, producing a spectrum a fraction of a second). As such it becomes ideal for parameter inference analyses.
A similar procedure was also used in \cite{Hart2020Sensitivity}, but here we extend the treatment to include second-order terms, thereby improving the precision and opening the path to study the effects of inhomogeneous recombination scenarios.

One of the most important features of \texttt{CRRfast} is that is it easily extendable. 
For instance, the inclusion of e.g., Dark Matter (DM) decay and annihilation \citep{Chluba2010Could} as well as EDE and VFC \citep{Hart:2022agu}, among others, would be straightforward. Furthermore, \texttt{CRRfast} has been made publicly available both as a stand-alone \texttt{python} code and as part of the cosmological Boltzmann solver \texttt{CLASS} \citep{Lesgourgues2011CosmicI, Blas2011Cosmic}, by default tightly interfaced with the parameter extraction code \texttt{MontePython} \citep{Audren2013Conservative, Brinckmann2018MontePython}. The inclusion of the CRR calculation into \texttt{CLASS} completes the set of \LCDM sources of SDs already implemented in \cite{Lucca2019Synergy}.
\texttt{CRRfast} therefore opens the door to many interesting developments and in the course of the manuscript several of them are discussed, setting up the stage for future dedicated analyses.

Using \texttt{CRRfast}, we can also study the effect of inhomogeneities in the recombination process on the CRR. In \LCDM, the effect is expected to be small, but it has not been quantified before. In addition, non-standard cosmologies, e.g., with inhomogeneous Big Bang Nucleosynthesis \citep[BBN --][]{Kajino1990, Jedamzik1994, Scherrer2021} or Primordial Magnetic Fields \citep[PMFs --][]{Jedamzik:2013gua, Jedamzik2020Relieving, Galli:2021mxk}, could cause more significant effects. This opens a new way to probe fluctuations in the Universe at redshifts $z=10^3-10^4$ even if these fluctuations are long gone or strongly altered today, as we illustrate here.

The manuscript is structured as follows. We begin in Sec.~\ref{sec: cosmo_dep} by briefly reviewing the cosmology dependence of the CRR. In Sec.~\ref{sec: num} we then present the numerical structure (Sec.~\ref{sec: Taylor_exp}) and implementation (Sec.~\ref{sec: gener}) of \texttt{CRRfast} both as a stand-alone emulator of \texttt{CosmoSpec} and as part of the \texttt{CLASS} code. In Sec.~\ref{sec: app} we proceed with an overview of the possible perspectives that might follow the development of \texttt{CRRfast}. Among others, in Secs.~\ref{sec: fore} and \ref{sec: exp} we present some examples of how \texttt{CRRfast} could be used to perform sensitivity forecasts and mission design studies. In Sec.~\ref{sec: var} we further extend the number of applications of \texttt{CRRfast} by focusing on the impact on the CRR of inhomogeneities in the recombination evolution. Specifically, after a general sensitivity forecast (Sec.~\ref{sec: detect}), we  estimate variance terms in \LCDM (Sec.~\ref{sec: LCDM}) and for models with PMFs (Sec.~\ref{sec: Hubble}). For the latter we also highlight the possible implications for the Hubble tension. We end the discussion in Sec.~\ref{sec: conc} with a summary and closing remarks.

%------------------------------------
\begin{figure*}
    \centering
    \includegraphics[width=\columnwidth]{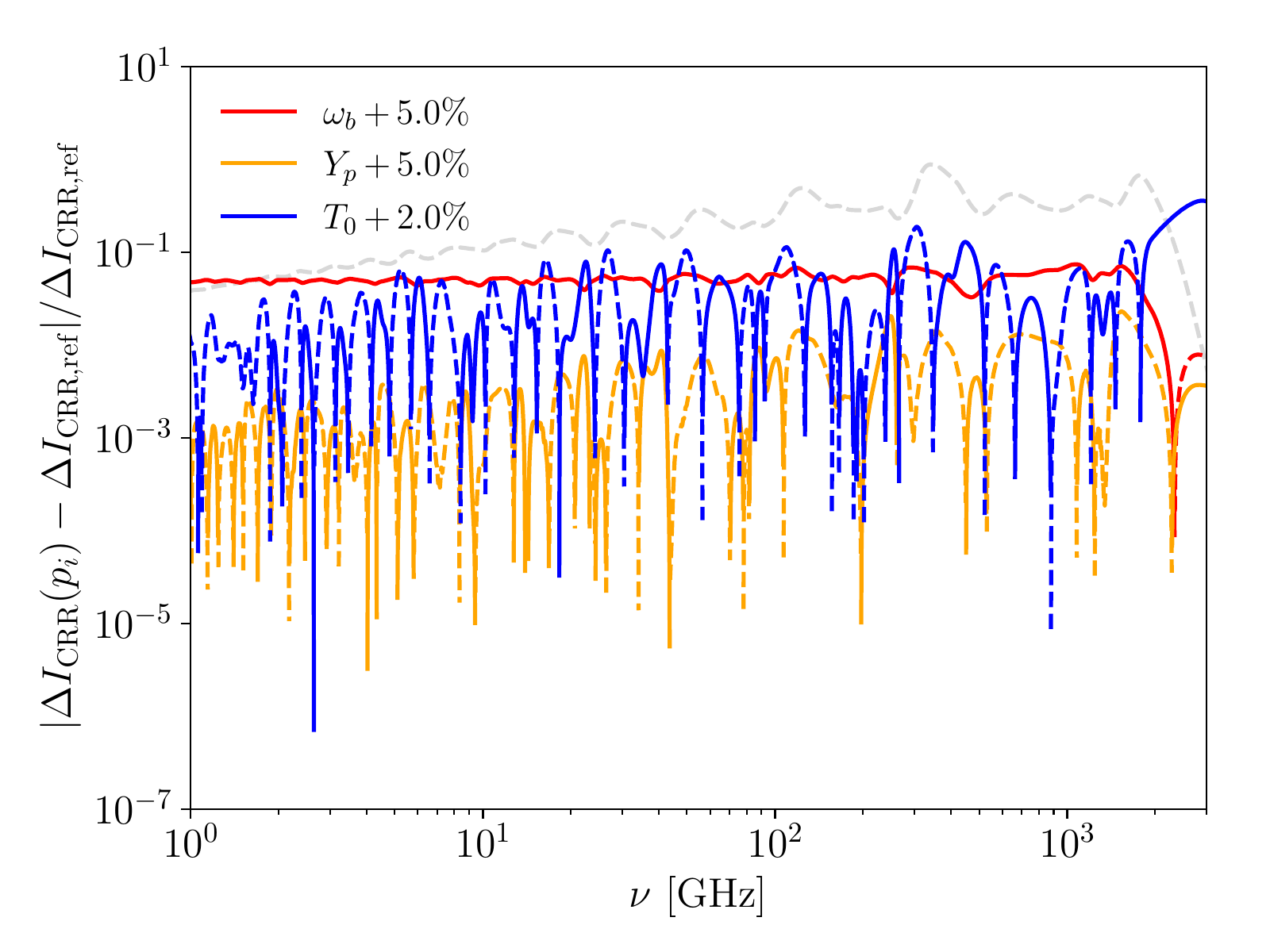}
   \includegraphics[width=\columnwidth]{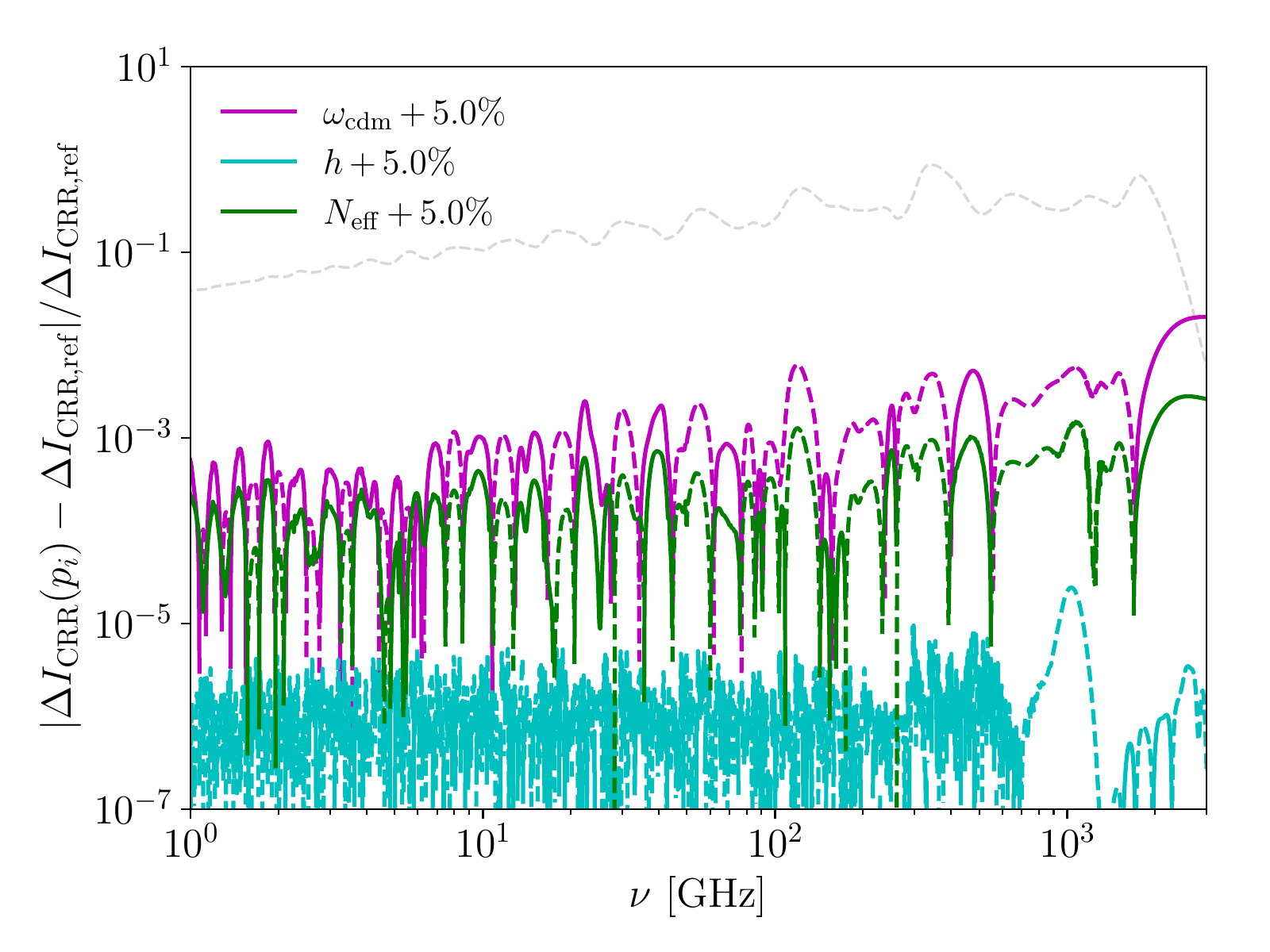}
   \\
\caption{Ratio of the CRR spectrum when varying the parameters $p_i$ listed in Eq.~\eqref{eq: params} ($\{\omb, Y_p, T_0 \}$ on the left and $\{\omega_{\rm cdm}, N_{\rm eff}, h\}$ on the right) with respect to the reference spectrum (showed in dashed gray in units of Jy/sr to facilitate comparisons). Solid and dashed lines represent positive and negative branches, respectively. For sake of completeness in the right panel we also show the case of the Hubble parameter $h$.}
\label{fig: DI_vs_ref}
\end{figure*}
%------------------------------------

\vspace{-3mm}
\section{The Cosmological Recombination Radiation}\label{sec: cosmo_dep}
%------------------------------------
We start by illustrating the dependence of the CRR spectrum on the main cosmological parameters that affect its shape. These are the baryon and DM energy densities, $\omb$ and $\omega_{\rm cdm}\,$, the dimensionless Hubble constant $h$, the helium to hydrogen abundance ratio $Y_p$, the CMB monopole temperature today $T_0$ and the number of effective relativistic degrees of freedom $N_{\rm eff}$. The remaining standard $\Lambda$CDM parameters, i.e., the reionization optical depth $\tau$ and the Primordial Power Spectrum (PPS) parameters $A_s$ and $n_s$, do not affect the recombination process directly and hence leave the CRR spectrum unaltered. We will therefore not consider them in this section.

The impact that the standard parameters have on the CRR spectrum has already been qualitatively and graphically explained in e.g., \citet{Chluba2008There, Hart2020Sensitivity}. In brief, there are three main ways via which these quantities affect the CRR. Firstly, $\omb$ and $Y_p$ define the hydrogen and helium abundances (as well as their ratio), which in turn determine how many photons are emitted during the recombination process and hence affect the amplitude of the spectrum. Secondly, $T_0$ determines the redshift of recombination and thereby the time at which the emissions occur. Therefore, a change in $T_0$ corresponds to a horizontal shift of the CRR spectrum (leaving the photons more or less time to redshift). Finally, $h$, $\omega_{\rm cdm}$ and $N_{\rm eff}$ primarily affect the ratio of the atomic time scales to the expansion rate of the universe around recombination and hence the photon escape rate. However, this effect is overall subdominant with respect to the others (in particular at low frequencies). 

These dependencies are graphically illustrated in Fig. \ref{fig: DI_vs_ref} \citep[see also Figs.~2-3 of][for a similar representation]{Hart2020Sensitivity}, where we vary in turn each one of the aforementioned parameters and show the ratio of the corresponding spectrum to the reference one (reported as dashed gray line to facilitate comparisons). As clear from the figure, $\omb$ and $T_0$ are the parameters with the largest impact on the CRR spectrum, the former imprinting an overall amplitude increase on the spectrum while the latter induces an oscillatory behaviour dictated by the horizontal offset explained above. The response to changes in $T_0$ in places is enhanced by roughly one order of magnitude due to the exponential dependence of the recombination time on this parameter \citep{Chluba2008There}. In comparison to the other parameters, this identifies $T_0$ as one of the variables that require a slightly higher precision for their representation.
The role of $Y_p$, $\omega_{\rm cdm}$ and $N_{\rm eff}$ is subdominant with respect to that of $\omb$ and $T_0$, although in particular for $Y_p$ the variations increase at high frequencies \citep[following the fact that the helium contribution also grows as a function of frequency, see also Sec. 2.2 of][for additional details]{Hart2020Sensitivity}. 

Among the shown parameters, the CRR is least sensitive to variations of $N_{\rm eff}$, as in the current setup it only enters in the expansion rate. Using the alternative parameterization $Y_{\rm p}=Y_{\rm p}(N_{\rm eff})$, however, modifies this picture as this leads to a direct dependence through variations of the helium contributions. Finally, $h$ has a negligible (and currently unobservable) impact on the CRR spectrum. We will therefore neglect its role in the following discussion, as also done in \cite{Hart2020Sensitivity}. Note, however, that this is due to our choice of parameter combinations. If we vary $\Omega_{\rm b}$ and $\Omega_{\rm cdm}$ instead of $\omega_{\rm b}=\Omega_{\rm b}\,h^2$ and $\omega_{\rm cdm}=\Omega_{\rm cdm}\,h^2$, a larger dependence is displayed.
We also point out that using a combination $\omega_{\rm b}/T_0^3$ and $\omega_{\rm cdm}/T_0^3$ as suggested by \citet{Ivanov2020} in connection with the recombination history does {\it not} yield a further reduction of the CRR dependencies (see Sec.~\ref{sec: Ivanov}).

One thus ends up with a set $p$ of five independent parameters,
%------------------------------------
\begin{align}\label{eq: params}
    p\equiv \{\omb,\, Y_p,\, T_0,\, \omega_{\rm cdm},\, N_{\rm eff}\}\,,
\end{align}
%------------------------------------
which can significantly affect the shape of the CRR spectrum. Henceforth (as well as in Fig. \ref{fig: DI_vs_ref}), we assume as reference values for the first three quantities the mean values reported by \cite{Aghanim2018PlanckVI} for the \Planck 2018+BAO combination (see Tab. 2 and Eq. (67b) of the reference), i.e., $\omb=0.02242$, $\omega_{\rm cdm}=0.11933$ and $N_{\rm eff}=2.99$. For $Y_p$ we will assume the value reported in Eq.~(82) of \cite{Aghanim2018PlanckVI}, $Y_p=0.2437$, which also takes into account additional information from BBN, while for $T_0$ we will assume the value reported by \cite{Fixsen2009Temperature}, $T_0=2.7255$~K, obtained from FIRAS data. The spectrum obtained with these choices will be referred to as reference CRR spectrum, $\Delta I_{\rm CRR, ref}$.

%------------------------------------
\begin{figure*}
    \centering
    \includegraphics[width=0.41\columnwidth]{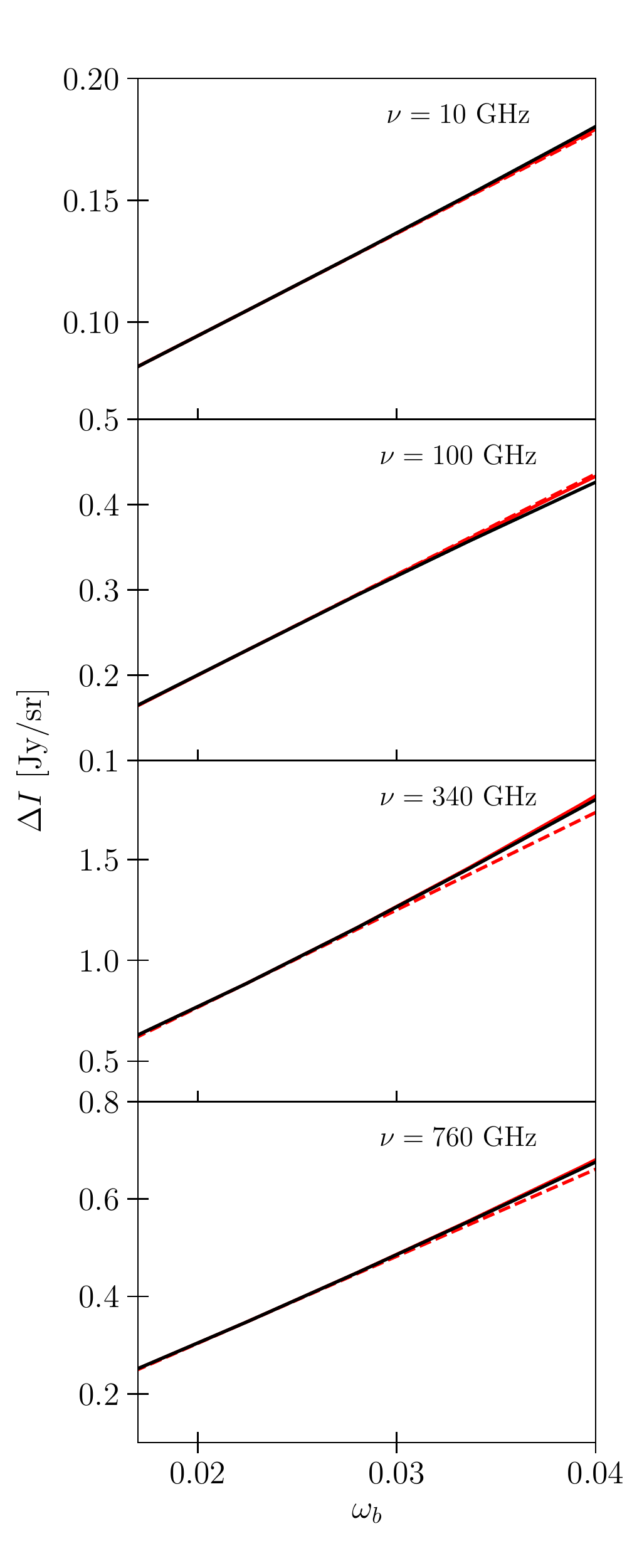}
    \includegraphics[width=0.41\columnwidth]{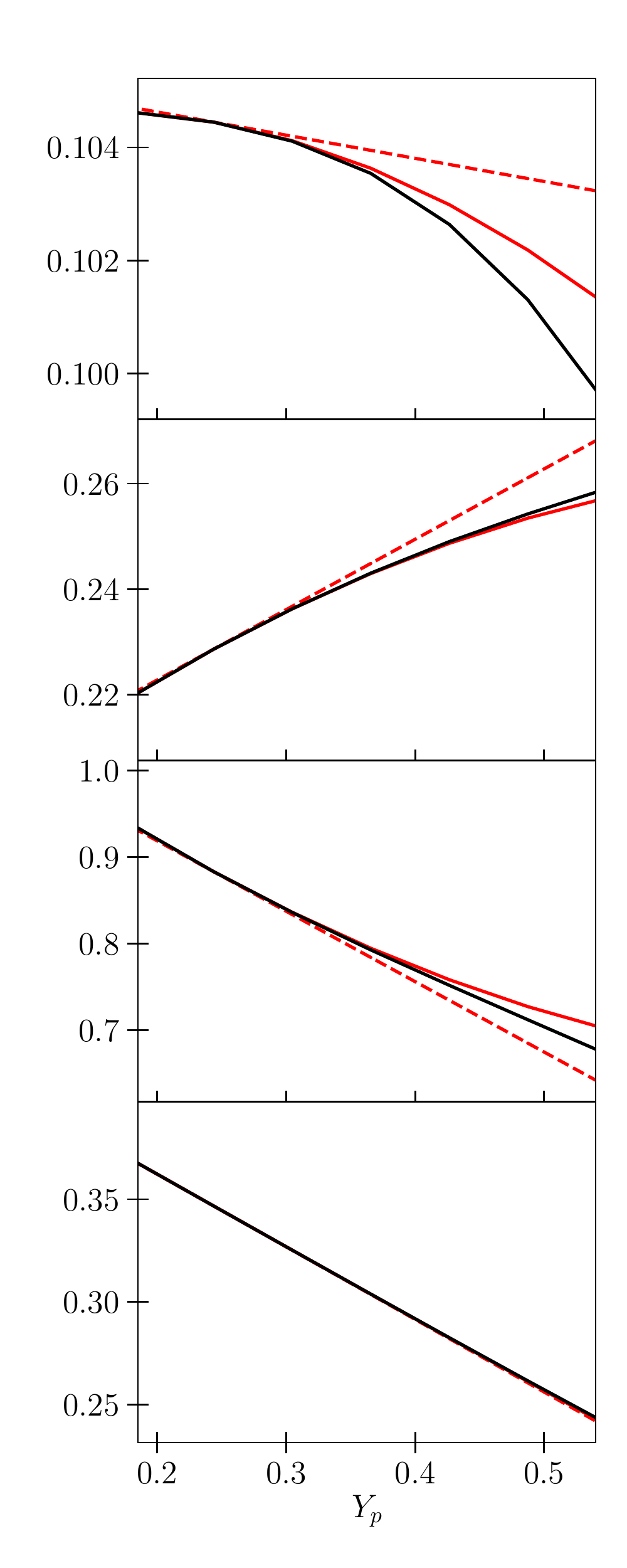}
    \includegraphics[width=0.41\columnwidth]{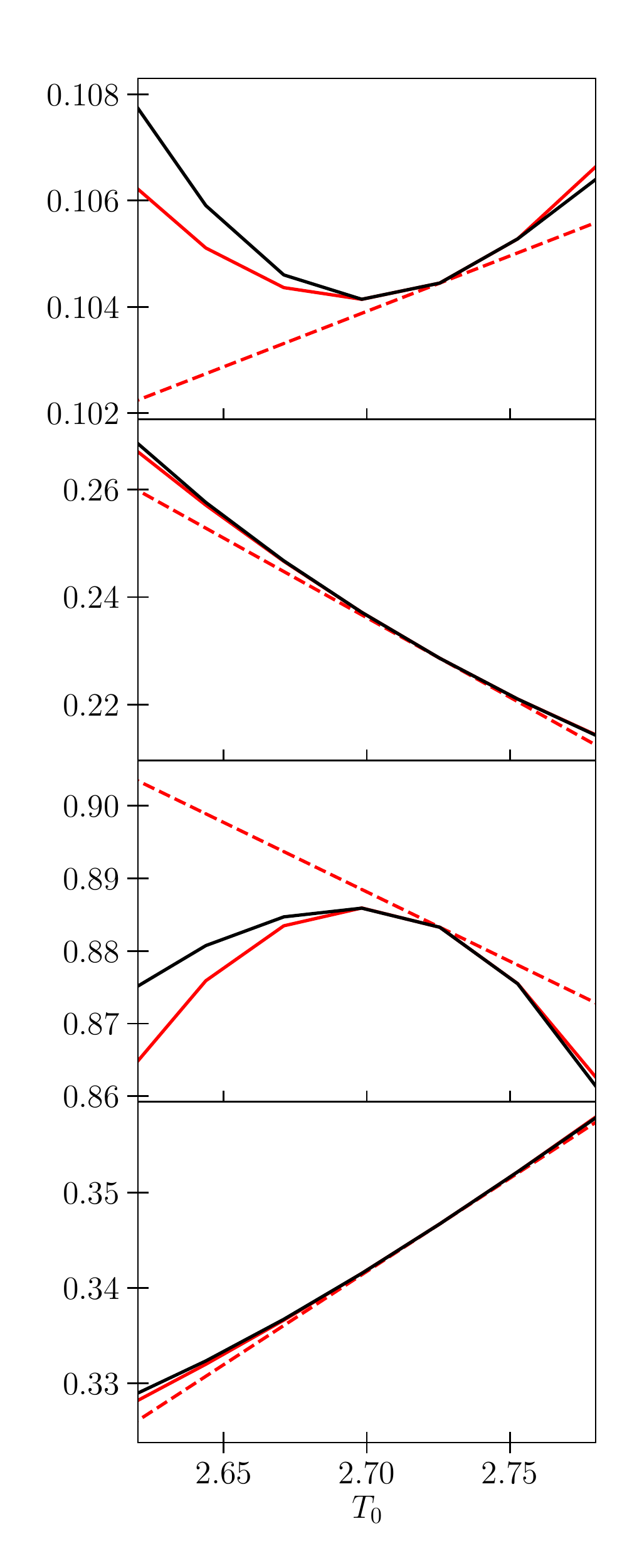}
    \includegraphics[width=0.41\columnwidth]{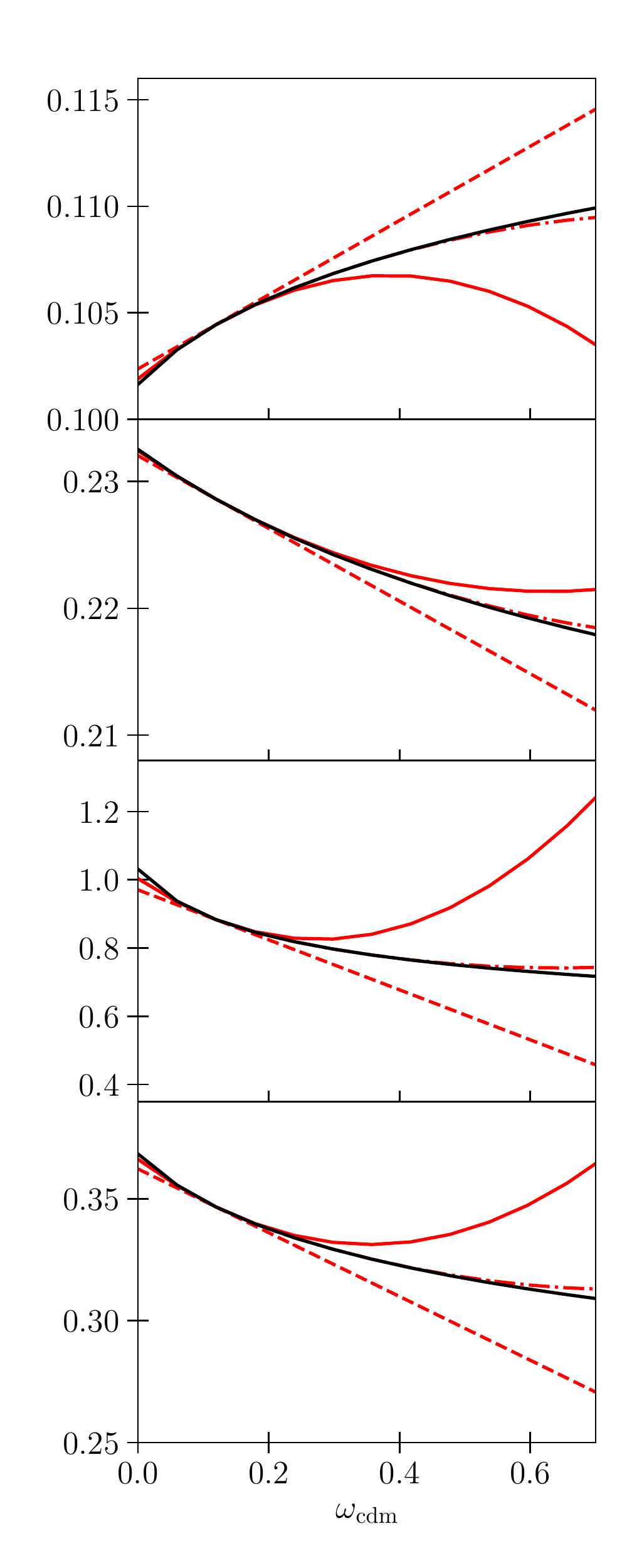}
    \includegraphics[width=0.41\columnwidth]{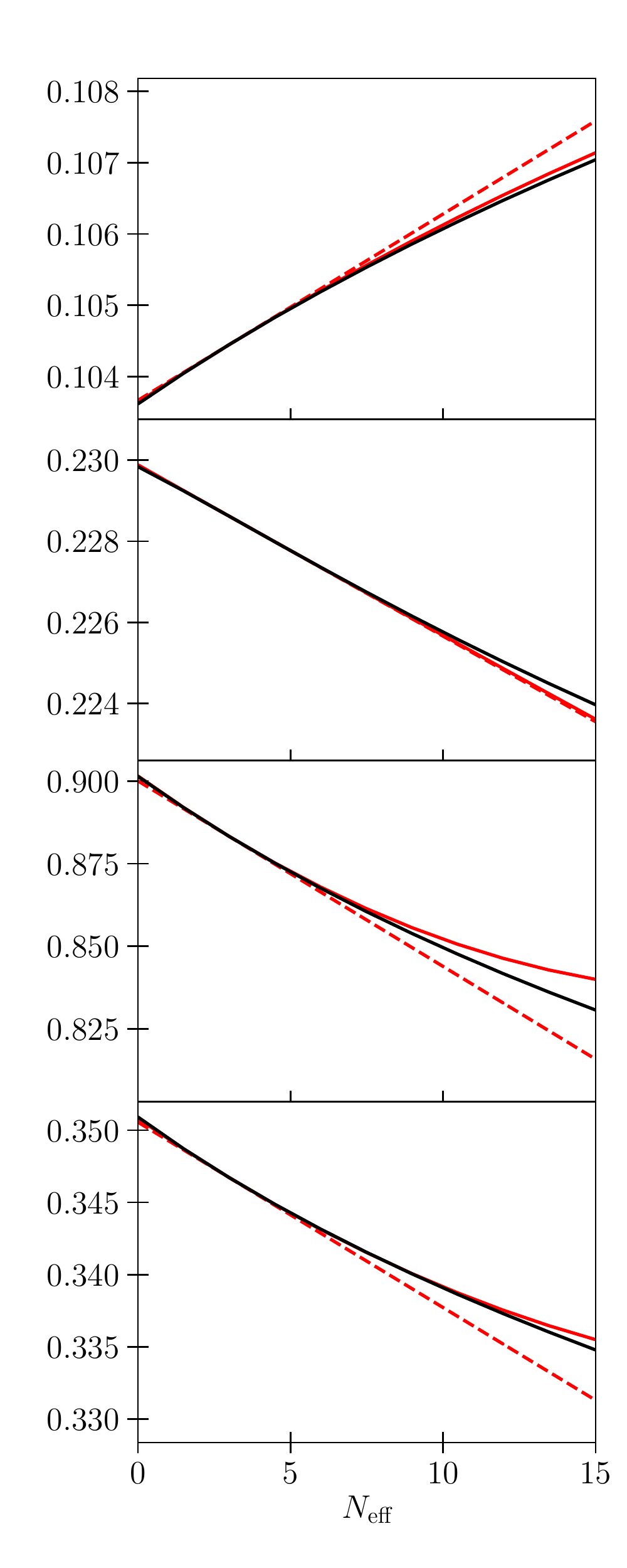}
    \\
    \vspace{-3mm}
    \caption{Fractional change of the CRR spectrum as a function of the parameters $p_i$ listed in Eq. \eqref{eq: params} evaluated at four arbitrary frequencies. In all subplots, the solid black lines are the reference curves computed with \CosmoSpec, the dashed and solid lines refer to the predictions of the Taylor expansion approximation using the first and second order coefficients, respectively. In the $\omega_{\rm cdm}$ case we also show as dashed-dotted line the prediction of the multi-pivotal approach.}
    \vspace{-3mm}
    \label{fig: DI_vs_params}
\end{figure*}
%------------------------------------

\vspace{-3mm}\section{\texttt{CRRfast}}\label{sec: num}
%------------------------------------
The goal of this section is to develop a simple and fast numerical setup able to accurately represent the CRR spectrum over a wide range of values of the underlying cosmological parameters, which we have listed and briefly presented in Sec.~\ref{sec: cosmo_dep}. For this we rely on the state-of-the-art \CosmoSpec code \citep{Chluba2016Cosmospec} and parameterise its cosmology dependence using a Taylor series expansion, whose details and limitations are discussed in Sec.~\ref{sec: Taylor_exp}. 
The resulting emulator, named \texttt{CRRfast}, is made publicly available both as a stand-alone \texttt{python} module and as part of the Boltzmann solver \texttt{CLASS}. 
The details of the respective numerical implementations are presented in Sec.~\ref{sec: gener}.

\subsection{Taylor expansion approximation}\label{sec: Taylor_exp}
%------------------------------------
To estimate more quantitatively how the CRR spectrum depends on the aforementioned parameters, in Fig.~\ref{fig: DI_vs_params} we display as solid black lines (computed using \CosmoSpec) how the CRR spectrum varies as a function of each parameter. We do so for four representative frequencies, $\nu=10, 100, 340$ (which corresponds to the Blamer-$\alpha$ peak) and $700$ GHz. The first important conclusion is that over the considered parameter range the scaling of the spectrum is almost perfectly linear with respect to $\omb$ for all considered frequencies, while for $Y_p$, $T_0$, $\omega_{\rm cdm}$ and $N_{\rm eff}$ second order corrections are necessary. The simplicity of these scalings allows us to rather accurately parameterize the dependence of the CRR spectrum with respect to these parameters as a second-order Taylor expansion around the fiducial. Concretely, we can express the CRR spectrum $\Delta I_{\rm CRR}$ as
%------------------------------------
\begin{align}\label{eq: taylor}
    \nonumber \Delta I_{\rm CRR} (\nu,p) & \simeq \Delta I_{\rm CRR, ref}(\nu)  + \sum_i \frac{\partial(\Delta I_{\rm CRR})}{\partial \ln p_i}
    \Bigg|_{\rm ref}\,
    \frac{\Delta p_i}{p^{\rm ref}_i} \\ & 
    \qquad\quad + \frac{1}{2}\sum_{i,j}\frac{\partial^2(\Delta I_{\rm CRR})}{\partial \ln p_i\partial \ln p_i}\Bigg|_{\rm ref}\,\frac{\Delta p_i}{p^{\rm ref}_i}\frac{\Delta p_j}{p^{\rm ref}_j}\,,
\end{align}
%------------------------------------
where $p_{i}$ can be any of the parameters in Eq.~\eqref{eq: params} and $\Delta p_{i}$ is the difference between $p_{i}$ and the chosen reference pivot, $p_{i}^{\rm ref}$. The first and second-order numerical derivatives can be easily computed as given in Appendix~\ref{app: details_Taylor}.

The predictions of the first and second order Taylor expansion are also shown in Fig.~\ref{fig: DI_vs_params} as dashed and solid red lines. As one can infer from the figure, the Taylor expansion at the second order allows for sub-percent precision at all the relevant frequencies and for sufficiently large variations, except for the $\omega_{\rm cdm}$ case. To address this issue, we employ a multi-pivot approach, introducing two extra pivot points at $\omega_{\rm cdm}=0.2178,0.3267$, which we find by minimizing the deviation from the reference set by \CosmoSpec and the number of required pivots. For a given value of $\omega_{\rm cdm}$, we then enforce our emulator to shift to the closest of the three fiducial values and employ the corresponding table of coefficients which we recalculate for every value of $\omega_{\rm cdm}$. The corresponding prediction of the CRR spectrum as function of $\omega_{\rm cdm}$ is shown in the respective panels of Fig. \ref{fig: DI_vs_params} as dashed-dotted line, which now overlaps sufficiently accurately with the \CosmoSpec prediction. 

In order to quantitatively assess the range of the variations within which this method can deliver sufficiently accurate results, we increasingly vary in turn every parameter $p$ up to the point where the relative difference between Taylor expansion and \CosmoSpec exceeds 1\%. We also ensure that the allowed range of variations covers the error bars predicted by \cite{Hart2020Sensitivity} for Voyage 2050, so as not to bias the forecasts for this mission (and those with better sensitivities) with numerical artefacts.
Based on this criterion, the multi-pivot Taylor expansion method is found to reliable within variations of the order of 50\% in the case of $\omb$ and $Y_p$, while much larger variations, of the order of 300\% and 400\%, are allowed for $N_{\rm eff}$ and $\omega_{\rm cdm}$, respectively. In terms of Signal-to-Noise Ratios (SNRs), these thresholds cover values as low as 2, 0.33 and 0.25, which means that the Taylor expansion approximation can be safely applied to sensitivity forecasts for Voyage 2050-like sensitivities and above \citep{Hart2020Sensitivity}. We also explicitly check that negative variations and the introduction of the cross-term contributions do not significantly change these conclusions. Moreover, deviations of the percent level in $T_0$ are also accurately reproduced (namely, up to 2\%), which are much larger than the current FIRAS uncertainty of 0.02\% \citep{Fixsen2009Temperature} and are therefore precise enough also for future missions.

\subsection{Numerical implementation}\label{sec: gener}
%------------------------------------

\subsubsection{Stand-alone implementation}
%------------------------------------
After having computed and tabulated the Taylor coefficients as described in the previous section, it becomes possible to implement the Taylor expansion approximation for the fast computation of the CRR spectrum in a simple \texttt{python} emulator, henceforth referred to as \texttt{CRRfast}. The code has been made publicly available\footnote{\url{https://github.com/luccamatteo/CRRfast.git}} together with the three tables of coefficients (one for each pivot value of~$\omega_{\rm cdm}$). As input it takes the array of values of the parameters listed in Eq.~\eqref{eq: params} within the validity range and outputs the CRR spectrum as a function of frequency.

For a \LCDM cosmology with typical \Planck values, the computation of the CRR with \CosmoSpec (assuming all relevant processes activated and precision settings) takes $\simeq 50$~s on a common laptop, a time that increases the larger the deviation from the standard cosmology. Although this is already significantly faster than in previous attempts to calculate the CRR \citep[see e.g., Sec. 4.1 of][]{Chluba2016Cosmospec}, with \texttt{CRRfast} this further reduces to
a fraction of a second, without significant loss of precision.

Overall, \texttt{CRRfast} complements the underlying \texttt{CosmoSpec} code, both in terms of scope and design. In fact, although \texttt{CRRfast} significantly speeds up computations and is thus better suited for time-demanding statistical analyses, \texttt{CosmoSpec} remains fundamental for the development of the physics of the CRR (including for instance effects beyond \LCDM) and for precision calculations. The development of the two codes should then proceed in parallel, thereby allowing to profit from the advantages of both.

As a final remark, we note that since \texttt{CRRfast} has been derived in the context of the $\Lambda$CDM model it might be inaccurate in exotic scenarios that affect the recombination history (see Sec.~\ref{sec: BLCDM}). This does not include, however, models only modifying e.g., the inflation history (see Sec.~\ref{sec: fore}).

\subsubsection{\texttt{CLASS} implementation}
%------------------------------------
Because of its speed, \texttt{CRRfast} lends itself to the implementation in Boltzmann solvers such as \texttt{CLASS} \citep{Lesgourgues2011CosmicI, Blas2011Cosmic}, designed for the fast computation of cosmological observables. In the particular case of \texttt{CLASS}, the inclusion of the CRR builds on the effort of previous works to implement CMB SDs in the code, which we briefly review below for context.

The \texttt{CLASS} implementation of SDs has been presented in great detail in \cite{Lucca2019Synergy} (whose notation we follow henceforth). Relying on the approximation scheme developed in \cite{Chluba2013Green}, it computes the final SD spectrum as an integral of the energy injection history, encoded in the heating rate $\dot{\mathcal{Q}}$, and a Green's function $G_{\rm th}$ that determines how much of the energy injection impacts the SD spectrum \citep[see Eq. (3.8) of][]{Lucca2019Synergy}. The Green's function, pre-computed using \CosmoTherm \citep{Chluba2011Evolution}, is further decomposed into energy branching ratios according to the prescription of \cite{Chluba2014Teasing}, which can then be used to compute the $y$ and $\mu$ parameters as well as the residual distortion coefficients. The heating rate includes in principle all sources of SDs that can be expressed in terms of such parameters (of note, this excludes the CRR). Here we will mainly focus on those predicted within the $\Lambda$CDM model (see e.g., \cite{Chluba2011Evolution, Chluba2016Which} as well as Sec. 2.4.3 of \cite{Lucca2019Synergy} for a review of these effects), and in particular on those that determine the primordial SD signal, i.e., the SD signal produced prior and around the epoch of recombination: the dissipation of acoustic waves \citep{Daly1991Spectral, Barrow199Primordial, Hu1994Power, Chluba2012CMB} and the adiabatic cooling of electrons and baryons \citep{Chluba2005Spectral, Chluba2011Evolution, Khatri2012Does}.\footnote{The contribution from the temperature differences of the CMB multipoles is also taken into account, but as a nuisance parameter at the level of the \texttt{MontePython} likelihood, as explained in Sec. 4 of \cite{Schoeneberg2020Constraining}.} 

The further inclusion of the CRR in this setup is one of the main novelties of this work. We achieve this by using the multi-pivot Taylor expansion approximation discussed in Sec. \ref{sec: Taylor_exp}. The tables are read by\footnote{For MCMC runs, the computation could be further sped up by pre-loading all tables once. However, at this point this does not cause a significant performance loss.} 
\texttt{CLASS} and the CRR spectrum is calculated depending on the input values of the relevant parameters discussed in Sec. \ref{sec: cosmo_dep}. The resulting spectrum is added on top of the contributions coming from the other effects.\footnote{Since the CRR spectrum cannot be expressed in terms of $y$ and $\mu$ parameters it is treated as completely independent of the other effects, similarly to how the Sunyaev-Zeldovich effect from low redshifts is also already accounted for in the code.} This completes the set of known $\Lambda$CDM sources of SDs accounted in the code.

As a final remark, we point out that the \texttt{CLASS} implementation of SDs has also been extended to the Markov Chain Monte Carlo (MCMC) code \texttt{MontePython} \citep{Audren2013Conservative, Brinckmann2018MontePython}, as extensively discussed in \cite{Lucca2019Synergy} (see Sec.~3.3 there) and \cite{Schoeneberg2020Constraining} (see Sec.~4 there). The former reference explains how to simulate the constraining power of various SD missions using mock likelihoods, while in the latter a number of galactic and extra-galactic foregrounds have been accounted for \citep[largely based on the previous work of][]{Abitbol2017Prospects}. This setup does not need to be modified to account for the inclusion of the CRR.\footnote{In principle, one should marginalize over possible variations of $T_0$ as also already done for temperature shifts \citep[see Sec. 4 of][]{Schoeneberg2020Constraining}. However, the sensitivity to $T_0$ for missions that could observe the CRR spectrum is so high that the impact of $T_0$ variations on the CRR are negligible \citep{Hart2020Sensitivity} and we can therefore safely neglect its marginalization.}

\section{Overview of possible applications}\label{sec: app}
%------------------------------------

The numerical tools developed in the previous section open many new doors for the theoretical and experimental studies of CMB SDs. In this section, we consider several of them. 

First of all, as highlighted in Sec. \ref{sec: fore}, thanks to the updated \texttt{CLASS}+\texttt{MontePython} implementation it is now possible to perform realistic sensitivity forecast for any SD mission including all \LCDM sources of SDs and a state-of-the-art foreground treatment. To further illustrate this aspect, in Sec. \ref{sec: exp} we discussed how the developed pipeline can be used to perform design studies for upcoming and future experimental setups targeting the CRR and CMB SDs more in general. Secs.~\ref{sec: BLCDM} and \ref{sec: time-dim} focus instead on possible extensions of \texttt{CRRfast} to include beyond-\LCDM cosmologies and to exploit more directly the time dependence of the CRR signal, respectively. Finally, Sec.~\ref{sec: Ivanov} shows that the CRR could be used to break the $\omb-T_0$ degeneracy that exists in CMB anisotropy analyses.

\subsection{Sensitivity forecasts for Voyage 2050 and Voyage 2050+}\label{sec: fore}
%------------------------------------
With the newly developed \texttt{CLASS} implementation, we can now forecast the sensitivity of any SD missions to the $\Lambda$CDM cosmology and some of its minimal extensions, taking into account all relevant contributions to the final spectrum. Following the analysis of \cite{Hart2020Sensitivity}, here we assume as baseline the Voyage 2050 mission \citep{Chluba2019Voyage}. As in the reference, we treat this mission as having the same frequency bands as Super-PIXIE~\citep{Kogut2019CMB}, but with all sensitivities improved by a factor of five (see App. \ref{app: details_exp} for the technical details). We also explore the constraining power of a mission with a sensitivity improved by a factor 10 with respect to Voyage 2050, henceforth referred to as Voyage 2050+. We implement the corresponding mock likelihoods in \texttt{MontePython} following the FIRAS example discussed in \cite{Lucca2019Synergy}. 

To start, we perform the same CRR-only forecasts as done in \cite{Hart2020Sensitivity} involving all parameter $p_i$ aside from $T_0$ (see Tab. 5 of the reference). To avoid introducing biases due to the different foreground treatments and since this exercise is only meant to test the CRR implementation, we only focus on the foreground-free case. We also fix the other $\Lambda$CDM parameters to the respective mean values found in \cite{Aghanim2018PlanckVI} for the \Planck 2018+BAO combination. As a result, we find perfect agreement between our forecasts and the results of \cite{Hart2020Sensitivity} for both Voyage 2050 and Voyage 2050+\,, confirming the validity of our numerical pipeline.

We can then move on to more comprehensive forecasts that involve all primordial sources of SDs. Given that within the $\Lambda$CDM model the dissipation of acoustic waves and the adiabatic cooling effects are mainly sensitive to the parameters $\omb$, $A_s$ and $n_s$ \citep[see e.g., Sec. 2.1 of ][]{Fu2020Unlocking} and the CRR to $\omb$ and $\omega_{\rm cdm}$ (see Sec.~\ref{sec: cosmo_dep}), we perform a scan in the parameter space $\{\omb,\, \omega_{\rm cdm},\, A_s,\, n_s \}$, which includes all $\Lambda$CDM parameters that SDs can constrain. To illustrate the principal sensitivity, we impose Gaussian priors with \Planck 2018 uncertainties \citep{Aghanim2018PlanckVI} on the remaining parameters, i.e., $h$ and $\tau_{\rm reio}$. 

We furthermore also perform a second scan extending the parameter space to $N_{\rm eff}$, i.e., with $\{\omb,\, \omega_{\rm cdm},\, A_s,\, n_s,\, N_{\rm eff} \}$. This parameter combination is particularly interesting since the impact of $N_{\rm eff}$ on the contribution from the dissipation of acoustic waves is perfectly degenerate with that of $A_s$ \citep[e.g., see Eq. (2.47) in][]{Lucca2019Synergy} and disentangling the two becomes possible with the additional information coming from the CRR (see Sec.~\ref{sec: cosmo_dep}). With the extended pipeline developed here we can thus explore the full constraining power of SDs and perform forecasts for this extended parameter space. Nevertheless, since the constraining power of Voyage 2050 on $N_{\rm eff}$ as an independent parameter is limited \citep{Hart2020Sensitivity}, we will perform this analysis only for Voyage 2050+. 

We remark that in the case of $N_{\rm eff}$ we do not account for the consistency relation between $N_{\rm eff}$ and $Y_p$ employed in e.g., \cite{Hart2020Sensitivity} (see Sec.~2.2.2 therein), so that the two parameters are treated independently. Including the consistency relation would further strengthen the bounds on $N_{\rm eff}$. In this sense, the following forecasted sensitivities to $N_{\rm eff}$ (and the parameters $N_{\rm eff}$ shares a degeneracy with) are to be taken as conservative, where $N_{\rm eff}$ is viewed as an independent parameter.

For both cases (with and without $N_{\rm eff}$) we marginalize over galactic and extra-galactic foregrounds following the \texttt{MontePython} implementation carried out in \cite{Schoeneberg2020Constraining}. We also marginalize over the presence of the late-time Sunyaev-Zeldovich (SZ) effect (at first order in the $y$ parameter, as done in the reference) as well as over the uncertainty on the experimental determination of the CMB monopole temperature.

Of course, these are just representative choices to highlight the constraining power of SDs in terms of the underlying cosmological parameters \citep[with previous estimates more focused on the SD signal itself, see e.g.,][]{Abitbol2017Prospects, Chluba2019Voyage}. Other beyond-$\Lambda$CDM parameters could be straightforwardly included as well, such as $Y_p$ and the running of the scalar spectral index\footnote{However, since they would only affect either one of the aforementioned sources of SDs (dissipation of acoustic waves and CRR), we do not expect any difference on the constraints from a combined analysis with respect to the dedicated forecasts already available in the literature (see e.g., \cite{Hart2020Sensitivity} and \cite{Fu2020Unlocking}, respectively).}, and many more models could be analysed once included in \texttt{CRRfast} (see Sec.~\ref{sec: BLCDM}). We leave this task for the future.

%------------------------------------
\begin{table}
	\def\arraystretch{1.2}
	\scalebox{1.0}{
	\begin{tabular}{l|ccccc} 
	    & $\omb$ & $\omega_{\rm cdm}$ & $A_s$ & $n_s$ & $N_{\rm eff}$ \\
	    \hline
        & \multicolumn{5}{c}{$1\sigma$} \\
	    \hline 
	    V2050 & $4.9\times10^{-3}$ & $0.19$ & $2.1$ & $0.14$ & $-$ \\
	    \hline
	    \multirow{2}{*}{V2050+} & $4.9\times10^{-4}$ & $0.019$ & $0.21$ & $0.014$ & $-$ \\
	    & $5.1\times10^{-4}$ & $0.022$& $0.33$ & $0.020$ & $2.5$ \\
	    \hline
        V2050+ w/o LFM & $6.7\times10^{-4}$ & $0.024$ & 0.23 & 0.015 & $-$ \\ 
	    \hline
	    \hline
        & \multicolumn{5}{c}{SNR} \\
	    \hline
	    V2050 & $4.9$ & $0.68$ & $1.0$ & $6.9$ & $-$ \\
	    \hline
	    \multirow{2}{*}{V2050+} & $49$ & $6.8$ & $10$ & $69$ & $-$ \\
	    & $45$ & $6.0$ & $6.4$ & $61$ & $1.2$ \\
	    \hline
        V2050+ w/o LFM & 34 & 5.0 & 9.2 & 64 & $-$ \\ 
        \hline
	\end{tabular} }
    \caption{Forecasted $1\sigma$ sensitivities (top) and SNRs with respect to the fiducial values of the cosmological parameters considered in our scans (bottom) for two representative SD missions (Voyage 2050, labeled V2050 in the table, and Voyage 2050+, labeled V2050+). The case of Voyage 2050+ is presented both with (as in the original proposal) and without the LFM.}
    \label{tab: SNRs}
\end{table}
%------------------------------------

%------------------------------------
\begin{figure}
    \centering
    \includegraphics[width=\columnwidth]{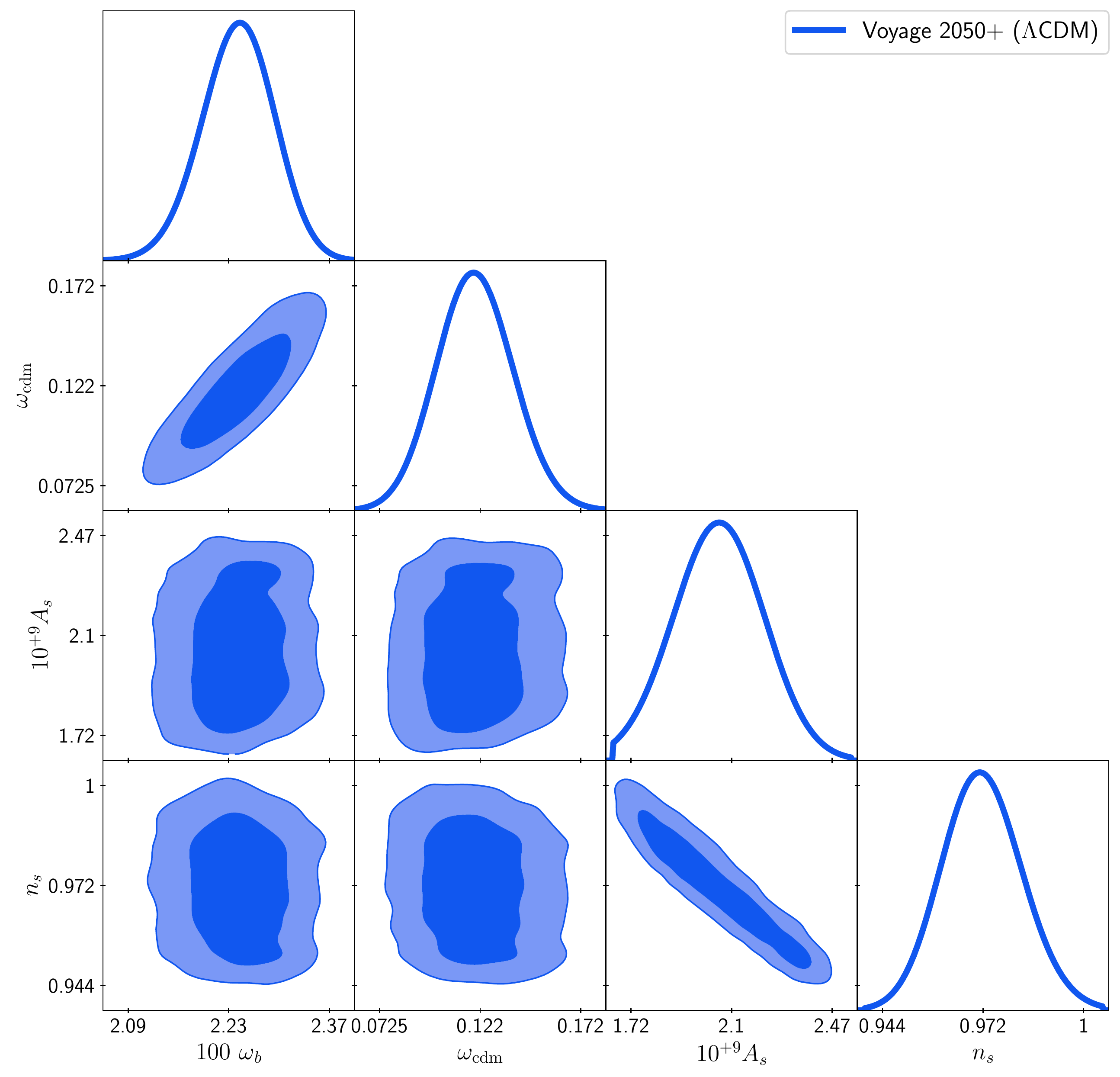}
    \caption{1D posteriors and 2D contours (at 68\% and 95\% CL) of the cosmological parameters that can be constrained with CMB SDs within the \LCDM model for the case of the Voyage 2050+ mission alone.}
    \label{fig: MCMC_res_1}
\end{figure}
%------------------------------------

The resulting $1\sigma$ sensitivities and SNRs are listed in Tab.~\ref{tab: SNRs}, while the full contours are graphically shown in Fig. \ref{fig: MCMC_res_1} for the \LCDM case of Voyage 2050+. As expected, the SNRs for $\omb$ and $\omega_{\rm cdm}$ are in excellent agreement with the results of \cite{Hart2020Sensitivity} in the presence of foregrounds (which further confirms the validity of the \texttt{CLASS}+\texttt{MontePython} implementation). This implies that the CRR dominates the constraining power for these parameters, in turn meaning that the CRR is the SD contribution most sensitive to the expansion history of the universe. 

In terms of $A_s$ and $n_s$, the MCMCs confirm the presence of the strong degeneracy between the two parameters already pointed out in \cite{Fu2020Unlocking} for high-sensitivity missions. The analysis carried out here, however, improves over the reference by including a more refined foreground modelling and a realistic experimental setup. It is also carried out solely in the context of CMB SDs, without the inclusion of complementary information from CMB anisotropy measurements, just for illustration. In this sense, the results presented in Tab. \ref{tab: SNRs} are original and show that even SDs alone can place stringent bounds on these quantities. 
 
Finally, the scan including $N_{\rm eff}$ shows a SNR very close to the one found in \cite{Hart2020Sensitivity}, suggesting that also in this case the CRR is more constraining than the dissipation of acoustic waves. The main difference between the \LCDM and \LCDM+$N_{\rm eff}$ scans is the significant worsening of the sensitivity to $A_s$ because of the degeneracy it shares with $N_{\rm eff}$ as explained above.

Overall, the forecasted error bars would not be competitive with the ones derived from e.g., CMB anisotropy data at face value, but such an accurate observation of the SD signal would still serve several important purposes even within the \LCDM model. First of all, as quantitatively shown here, SDs would be able to constrain four \citep[and, optimistically, five including $H_0$ as a byproduct, see][]{Abitbol2019Measuring} of the six \LCDM parameters at the same time with statistically significant SNRs. The ability to constrain so many \LCDM parameters in absence of any other cosmological information is second only to CMB anisotropy data. As such, CMB SDs would be able to deliver a totally independent and complete picture of the expansion and thermal history of the early universe.

Moreover, the fact that CMB SD sensitivities are not competitive with CMB anisotropy constraints at face value does not mean that the former would not be able to improve upon the latter once combined. In fact, as shown in \cite{Hart2020Sensitivity} and \cite{Fu2020Unlocking} in the context of the CRR and $\mu$ distortions, respectively, the combination of CMB SDs with \Planck and up-coming CMB anisotropy missions would deliver more stringent error bars than in the CMB anisotropy-only cases. 
Furthermore, the constraints on the PPS parameters would apply to very different scales than the ones probed by CMB anisotropy experiments \citep[e.g.,][]{Chluba2012Inflaton, Khatri2012Mixing, Chluba2015Features, Schoeneberg2020Constraining}. Given the extended level arm the SDs offer at small scales the combined determination of $n_s$ would improve upon the CMB anisotropy-only case by a factor of $2-3$ even with respect to future CMB anisotropy missions.\footnote{In this context, an important remark is that the relative improvement of CMB SDs with respect to CMB anisotropies is mainly due to the pivot scale used as default to define the PPS parameters. If instead of $k_*=0.05$ Mpc$^{-1}$ one used $k_*=50$ Mpc$^{-1}$ the relative constraining power of the two CMB probes would tilt towards SDs.} Furthermore, discrepancies of the order of $10^{-3}$ in $\omb$, which would be testable with Voyage 2050+ sensitivities, are currently being reported by various CMB anisotropy experiments \citep[see Tab. 4 of][]{ACT:2020gnv}. Although upcoming CMB anisotropy surveys will most likely be able to shed light on the origin of this inconsistency on a shorter timescale, this still underlines the fact that the aforementioned SD missions would start to test a sensitive region of parameter space in a CMB-anisotropy-independent way and at similar cosmological times even within \LCDM.

Finally, a percent understanding of the galactic and extra-galactic foregrounds has been shown to be able to improve the sensitivity to the $\mu$ signal by about one order of magnitude \citep{Abitbol2017Prospects}. Achieving this degree of precision and beyond on Voyage 2050 timescales is realistic and would further improve the forecasted sensitivities. A similar argument could be also be made for the detection of the CRR (a task that we leave for future work).

\vspace{-3mm}
\subsection{Experimental design}\label{sec: exp}
%------------------------------------
The first systematic analysis of the relation between the experimental characteristics of a given SD mission and its sensitivity to cosmological parameters has been performed in \cite{Fu2020Unlocking}. It considered several experimental environments and sensitivities, and focused on the interplay between CMB anisotropy and SD missions. Although presenting useful order-of-magnitude estimates for a variety of scenarios, the numerical pipeline had, however, some important limitations such as the exclusion of the CRR and of (extra-)galactic foregrounds. With this and the work carried out in \cite{Schoeneberg2020Constraining}, these missing pieces have been accounted for and it is now possible to perform state-of-the-art forecasts with the \texttt{CLASS}+\texttt{MontePython} pipeline for any SD mission design.

To showcase the potential of these tools, here we focus on the role of the main novelty introduced to upgrade the PIXIE mission \citep{Kogut2011Primordial} to its advanced Super-PIXIE version \citep{Kogut2019CMB}, which is the basis for the Voyage 2050 proposal \citep{Chluba2019Voyage}, namely the Low-Frequency Module (LFM). As displayed in e.g., Fig. 9 of \cite{Chluba2019Voyage}, the LFM is designed to span the frequency range between 10 and 40 GHz with a 2.5~GHz bin size and sensitivity a factor of about $4-5$ better than in the mid- and high-frequency modules. The main reason for the introduction of this additional module is that it would provide an improved sensitivity where the relative difference between the late-time SZ signal and the primordial $\mu$ signal is expected to be the largest, thereby increasing the detectability of the latter \citep{Abitbol2017Prospects}.

A precise forecast assessing the actual improvement of the LFM in the specific design of Voyage 2050 has not been performed so far. It is in particular unclear, \textit{a priori}, if the improvement will also benefit the detection of the CRR or not. For this reason, here we repeat the \LCDM run discussed in the previous section in the context of the Voyage 2050+ mission, but without the inclusion of the LFM. Since the detection of the CRR completely determines $\omb$ and $\omega_{\rm cdm}$, while the information carried by the $\mu$ signal translates into bounds on $A_s$ and $n_s$, studying how the respective constraints change will be indicative of how important the LFM is for the design envisioned in the Voyage 2050 concept.

The results are summarized in Tab. \ref{tab: SNRs}. As it turns out, the presence of the LFM in the Voyage 2050 frequency array does not seem to yield a significant improvement with respect to the setup without it. 
This is primarily because most of the signal-to-noise gains from the CRR are expected from higher frequencies \citep[e.g.,][]{Vince2015Detecting}, suggesting that for the CRR the role of the LFM is not as crucial
Although this finding deserves a dedicated analyses with e.g., varying frequency arrays and sensitivities, it still highlights the gain to be had using the pipeline developed here for cost-reward analyses of this type.
In particular, in terms of the CRR it may also be possible to significantly improve the performance of the MF and HF modules with more sophisticated processing of the frequency information.

%------------------
\begin{figure*}
    \centering
    \includegraphics[width=0.48\textwidth]{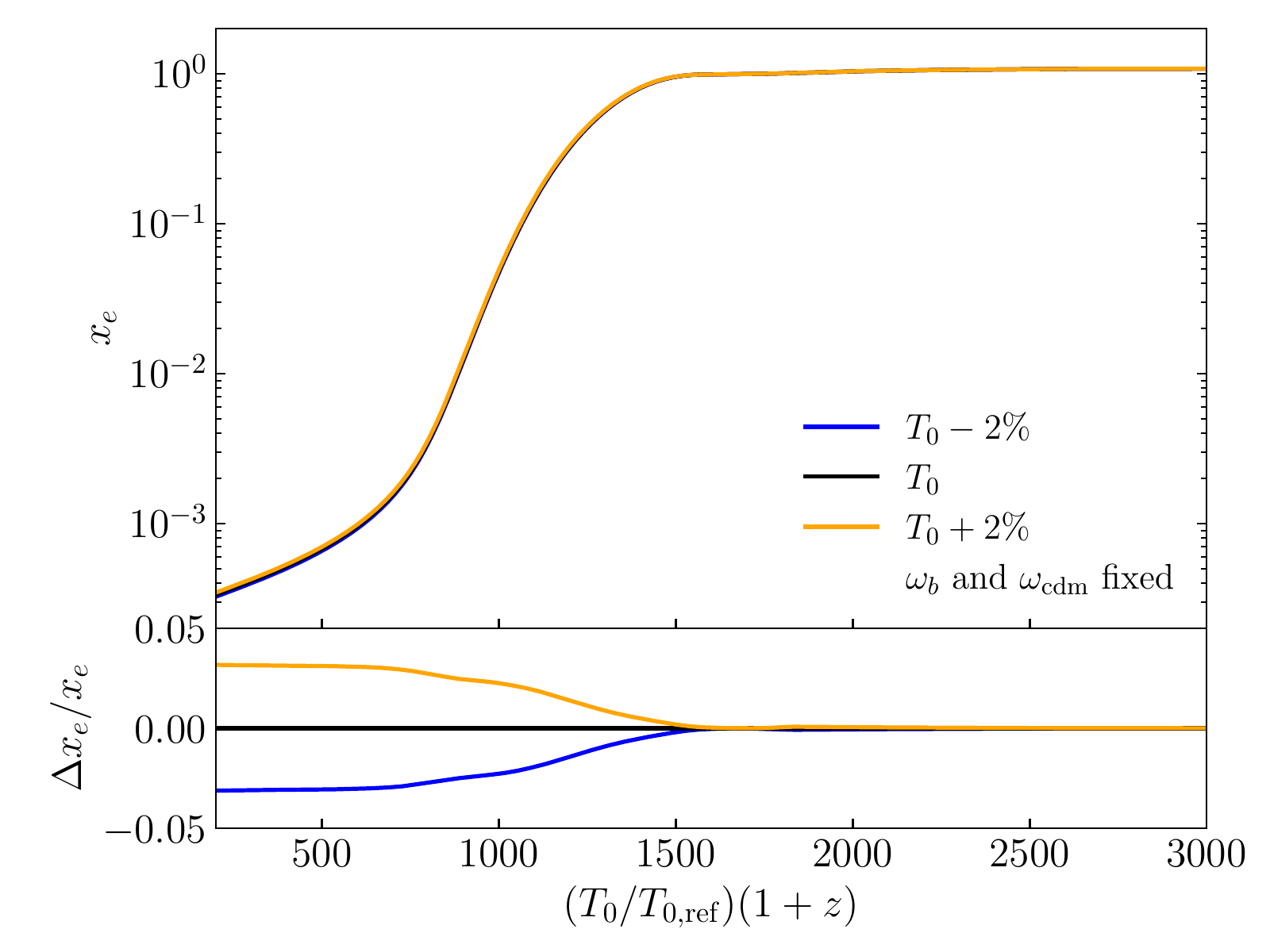}
    \includegraphics[width=0.48\textwidth]{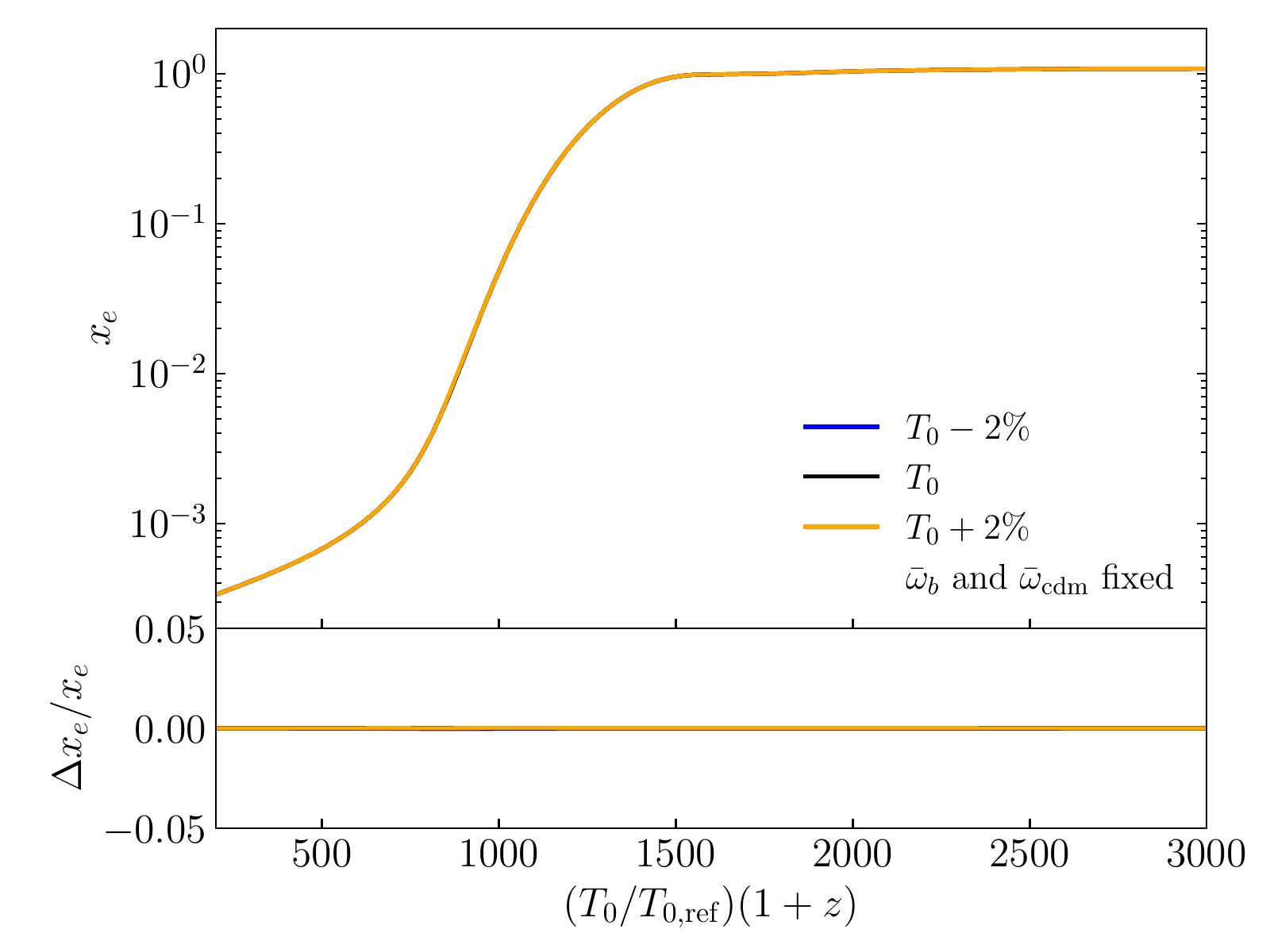}
    \caption{Absolute (top) and relative (bottom) evolution of the free electron fraction for different values of the CMB temperature today. In the left panel the values of $\omb$ and $\omega_{\rm cdm}$ are kept fixed, while in the right panel the values of the parameter combinations $\bar{\omega}_i=\omega_i/T_0^3$ are kept fixed, where $i=b,$cdm. The figure reproduces Fig. 1 of \citet{Ivanov2020}, with the difference that here the $x$-axes are consistently chosen between the two sub-plots.}
    \label{fig: Ivanov}
\end{figure*}
%------------------

Beyond these interesting conclusions, we also point out that the utility of the current setup is not limited to design studies, but can also be employed in the context of on-going SD missions. A prime example would be the aforementioned APSERa experiment, designed to target the $3-6$~GHz frequency range with nK sensitivity, forecasted to be enough to detect the CRR signal with high significance \citep{SathyanarayanaRao:2015vgh,SathyanarayanaRao:2016ope}.
These types of measurements would nicely complement space-based approaches, essentially providing the means to test the physics of recombination in yet another frequency band.

\vspace{-3mm}
\subsection{Beyond-\LCDM extensions}\label{sec: BLCDM}
%------------------
The current implementation of \texttt{CRRfast} can be extended in several directions in the future. One of the main avenues to pursue is the extension of the code to all known beyond-\LCDM models that might affect the CRR. Examples that are already implemented in \CosmoSpec include, among others, annihilating and decaying DM, EDE, VFC and PMFs (see following section). Their inclusion in \texttt{CRRfast} would be straightforward, as it would simply require to derive the respective Taylor coefficients by varying the underlying non-standard parameters. Other scenarios might involve DM-photon scatterings, Primordial Black Hole (PBH) evaporation and several other exotic sources of heating. In order to include such cases, the models would first need to be accounted for at the level of \CosmoSpec, to be then treated as the aforementioned examples. 

The inclusion of each of these models in the \texttt{CRRfast} would have direct applications. For instance, although the impact of EDE on the CRR spectrum has already been studied in \cite{Hart:2022agu}, its implementation in \texttt{CRRfast}, once propagated to \texttt{CLASS}, would in turn allow to explore the role of the CRR in combination with other relevant data sets. Moreover, the constraining power of the CRR with respect to the energy injection scenarios mentioned above has not been considered systematically in the literature \citep[see][for a preliminary study]{Chluba2010Could} and a dedicated analysis would be of major interest. It would be in particular useful to know if, for a mission sufficiently accurate to observe the CRR, the bounds imposed by the CRR would supersede the standard ones derived from the $\mu$ distortions. 
This would also require a treatment of uncompensated atomic transitions in the presence of non-equilibrium CMB radiation \citep{Chluba2009Pre}, which indeed requires a combination with {\tt CosmoTherm}.
We leave a an exploration of these directions to the future.

\subsection{Access to the time dimension}\label{sec: time-dim}
%------------------
As mentioned in the previous sections, the emulator could be expanded (in a straightforward way) to include other models already implemented in \CosmoSpec such as decaying DM and EDE \citep{Chluba2010Could, Hart:2022agu}. In addition, this analysis sets the stage for another interesting perspective: the current Taylor expansion approximation only captures the total CRR spectrum, but it could be easily split in its dependence on the three different recombination eras \citep[e.g.,][]{Sunyaev2009Signals}. Having access to this type of information would allow to set limits on quantities such as e.g., the monopole temperature $T_0$ as a function of time around the respective epochs of recombination. No other cosmological probe would be able to deliver similar constraints.

For instance, given that a percent change in $T_0$ (i.e., of the order of a few mK) leads to a $\simeq 1-10\%$ change in the CRR spectrum (see Fig.~\ref{fig: DI_vs_ref}), a Voyage 2050 mission sensitive to variations of the order of 10\% \citep{Chluba2019Voyage} might be able to deliver $\mathcal{O}$(mK) constraints on $T_0$ at the time of hydrogen recombination.\footnote{Note that this is very different from the much more precise bounds \citep[of $\mathcal{O}$(nK), see][]{Chluba2019Voyage} that a Voyage 2050 mission would deliver based on the precise observation of the BB spectrum, since the latter would be determining $T_0$ today.} Because of the secondary helium contribution, the same type of precision cannot be expected for $T_0$ at the epoch of helium recombination, but even looser constraints of the order of $10-100$~mK would be exceptionally useful to test the time-dependence of the temperature-redshift relation and would be in any case the first of their kind. We leave a dedicated quantitative analysis for future work.

A similar discussion also applies to other quantities and models that affect the CRR in a time-dependent way. For instance, the analysis presented in \cite{Hart:2022agu} clearly shows that an accurate observation of the CRR would provide a new time-sensitive constraints of the evolution of the EDE fluid precisely around the time it becomes dynamic. This would allow us to shed directly light on the expansion history in the (pre-)recombination era. In addition, the variation of fundamental constants could be probed.

%------------------------------------
\begin{figure*}
    \centering
    \includegraphics[width=0.48\textwidth]{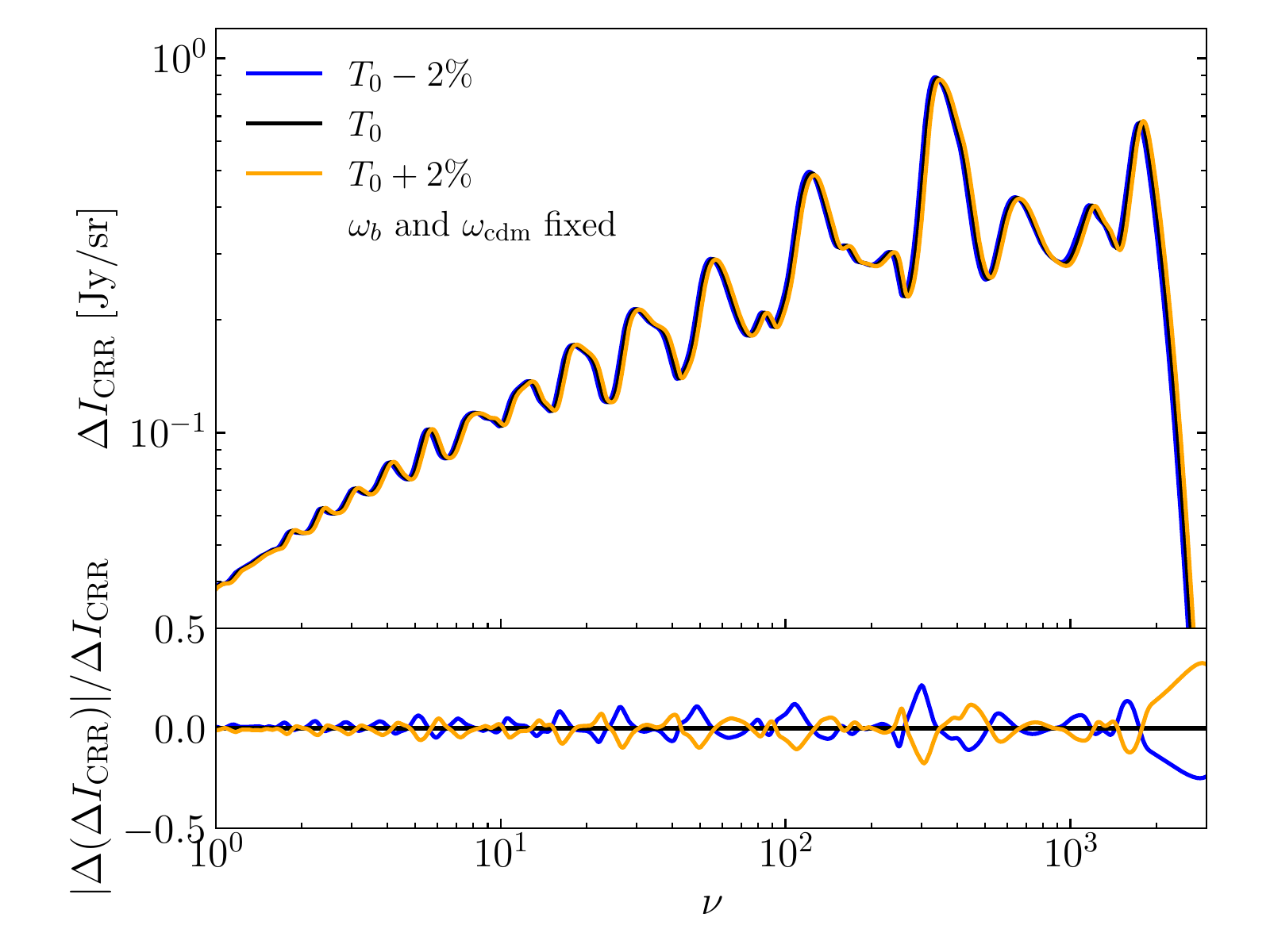}
    \includegraphics[width=0.48\textwidth]{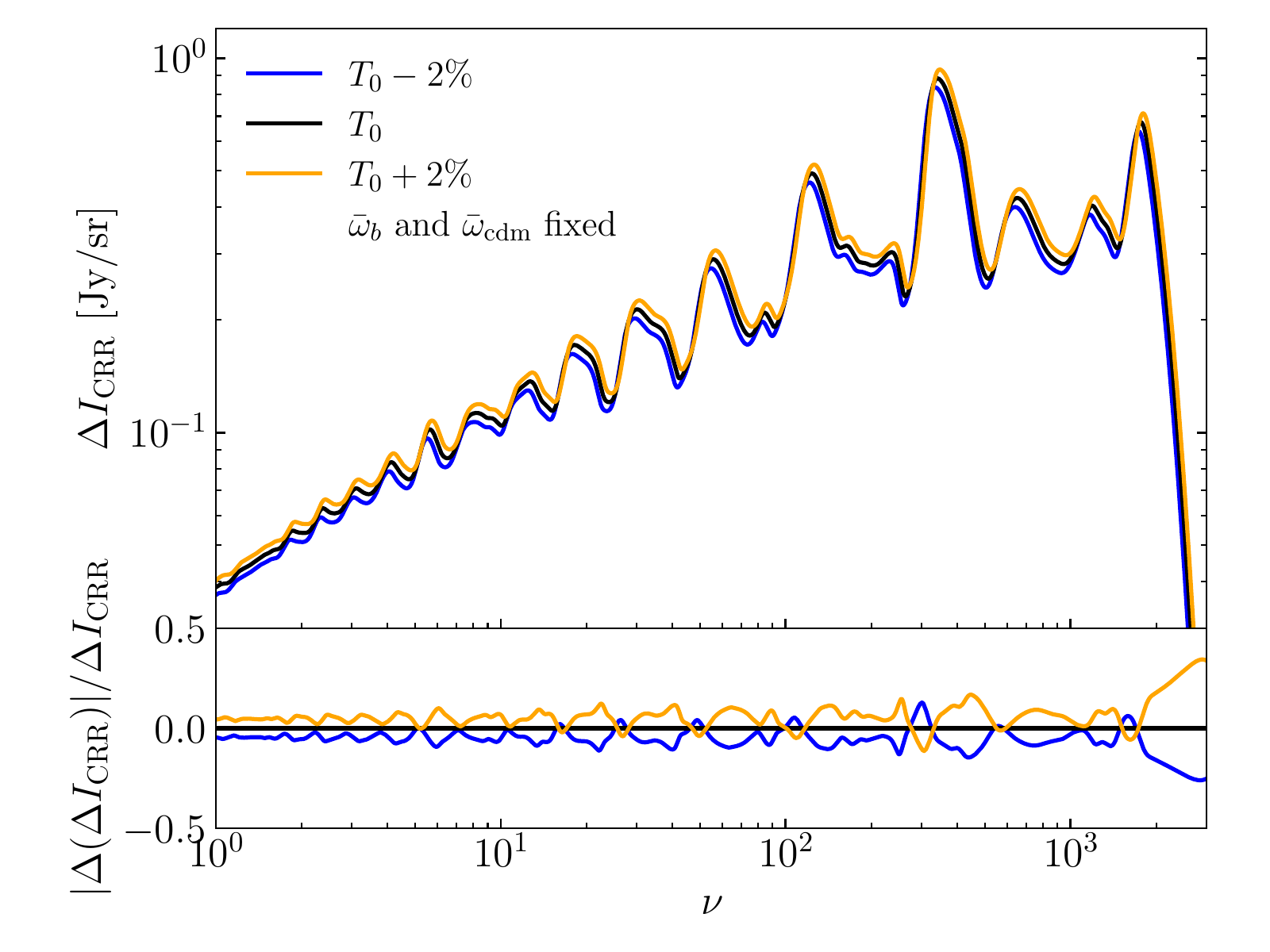}
    \\
    \includegraphics[width=0.48\textwidth]{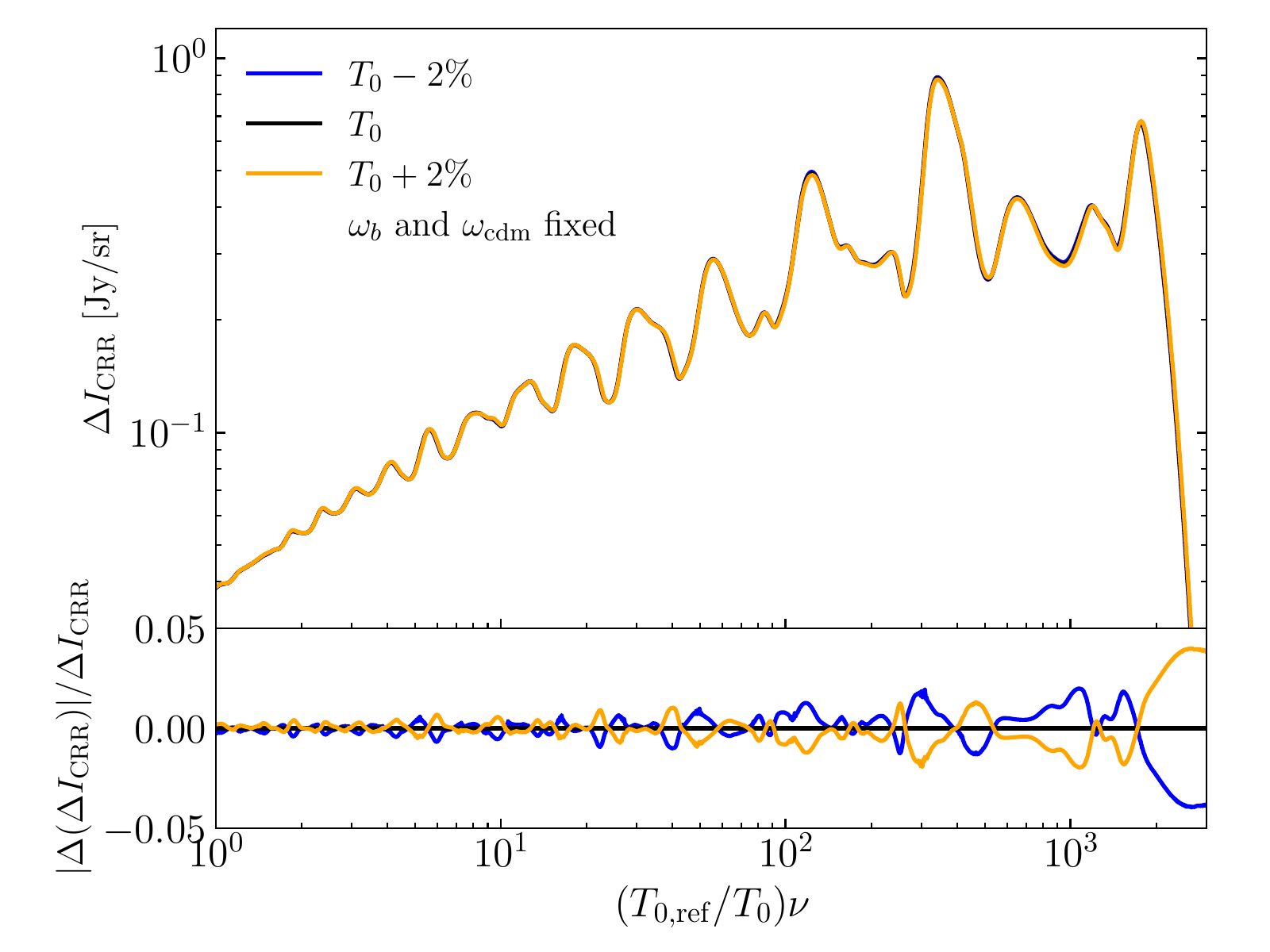}
    \includegraphics[width=0.48\textwidth]{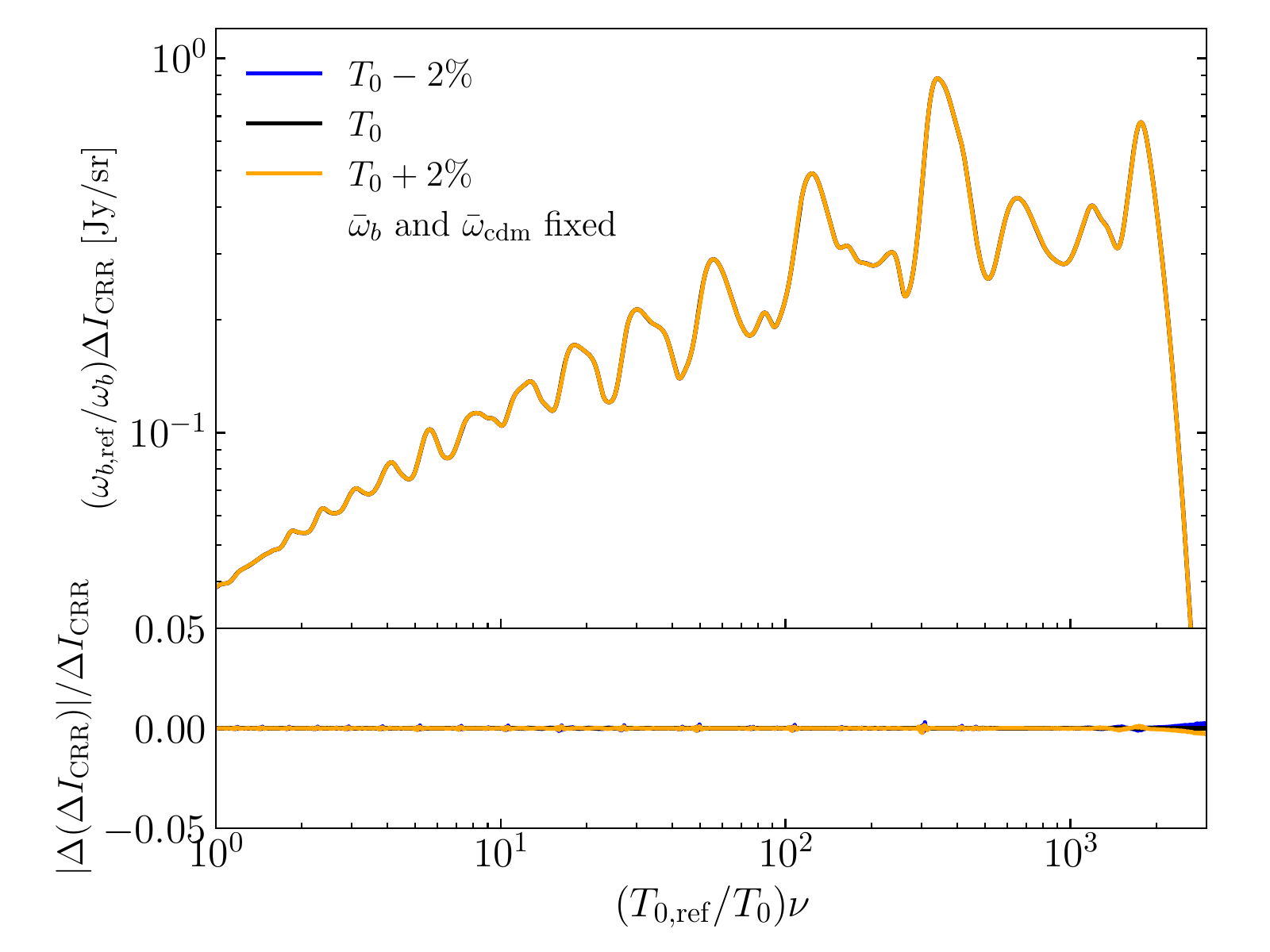}
    \\[-3mm]
    \caption{Same as in Fig.~\ref{fig: Ivanov}, but for the CRR spectrum. \textit{Top panels:} CRR spectra with unscaled axes. \textit{Bottom panels:} In addition to changing the frequency scale we respectively have to multiply the CRR by the ratios $\omb/\omega_{b, \rm ref}$ to achieve approximate invariance (in the right panel the rescaling factor is equal to unity and therefore omitted). Note the one order of magnitude difference in the lower panels of the top and bottom figures. While this can be useful in compressing the information for the emulator, it also shows that the CRR transformation still depends on $T_0$ and the values of $\omb$ and $\omega_{\rm cdm}$.}
    \vspace{-3mm}
    \label{fig: Ivanov_2}
\end{figure*}
%------------------------------------

\subsection{Breaking the degeneracy between $\omb T_0^3$}
\label{sec: Ivanov}
%------------------
The last application we highlight involves the cosmological degeneracy between the baryon and DM energy densities $\omb$ and $\omega_{\rm cdm}$ and the CMB monopole temperature $T_0$ \citep[see e.g.,][for a recent discussion]{Ivanov2020}. In fact, as shown in Fig.~\ref{fig: Ivanov} \citep[which reproduces Fig.~1 of][]{Ivanov2020}, while a change in $T_0$ would \textit{per se} affect the free electron fraction, $x_e$, this is only true as long as the values of $\omb$ and $\omega_{\rm cdm}$ are kept constant. If instead one fixes the parameter combination $\bar{\omega}_i=\omega_i/T_0^3$, where $i=b,$cdm, $x_e$ would remain unaffected\footnote{This is analogous to how the CRR becomes insensitive to $h$ once the parameter combinations $\omb$ and $\omega_{\rm cdm}$ are employed instead of $\Omega_{\rm b}$ and $\Omega_{\rm cdm}$ (see Sec.~\ref{sec: cosmo_dep}).} when the redshift is appropriately rescaled (or equivalently, when the ionization history is given as a function of the temperature). A similar scaling also propagates to the CMB anisotropy power spectra \citep[see Sec.~3.5 of][although it would not apply to e.g., the matter and CMB lensing power spectra]{Ivanov2020}. 
This also implies that the CMB anisotropies alone can only weakly constrain $T_0$ unless additional data from BAO is added \citep{Ade2015PlanckXIII, Ivanov2020}.

Nevertheless, as argued in Sec.~\ref{sec: cosmo_dep} and explicitly illustrated in Fig.~\ref{fig: Ivanov_2}, the degeneracy can be broken with the CRR. The reason for this is that, at least for the baryon energy density, the effect of variations of $\omb$ and $T_0$ on $\Delta I_{\rm CRR}$ is orthogonal, since the former moves the spectrum vertically while the latter moves it horizontally. 
This simply follows from the fact that $\Delta I_{\rm CRR}\propto \omb$ and that redshifting leaves the distortion spectrum constant at $x=h\nu/k T_0(1+z)$.
As shown in the left panels of Fig.~\ref{fig: Ivanov_2}, varying $T_0$ for fixed $\omb$ and $\omega_{\rm cdm}$, and then scaling the frequency appropriately reduces the variations in the spectrum significantly. When instead varying $T_0$ for fixed $\bar{\omega}_b$ and $\bar{\omega}_{\rm cdm}$, by scaling the amplitude of the CRR by $\omega_{b, \rm ref}/\omb=(T_{0, \rm ref}/T_0)^3$ in addition to the frequency one can collapse the spectra to within the numerical precision of the treatment (see right panels). While this may allow a compression of the representation, it does not eliminate any degree of freedom, still requiring three parameters to perform the mapping.

Overall, the CRR thus does not suffer from the same degeneracy between the energy densities and $T_0$ present when analysing the CMB anisotropy power spectra \citep[see e.g., Fig.~5 of][]{Ivanov2020}. This means that $T_0$ (in fact evaluated at the time of recombination) can actually be constrained with the CRR, justifying the suggestion made in the previous section. Moreover, this result implies that an observation of the CRR could in principle be combined with CMB anisotropy data to yield a precise CMB-only determination of $H_0$. Referring to Fig.~5 of \cite{Ivanov2020}, the addition of the CRR to \Planck would help breaking the degeneracy between $T_0$ and $H_0$ (since the CRR would be able to constrain $T_0$), thereby tightening the constraints on the latter. This is similar to the role that Baryon Acoustic Oscillation (BAO) data plays in the problem, with the difference that for the CRR the determination of $H_0$ would solely rely on CMB information and would therefore represent an important new consistency check.

Finally, we remark that the remapping of the ionization history and the CRR performed by using $T_0$, $\bar{\omega}_b$ and $\bar{\omega}_{\rm cdm}$ does not represent an {\it exact} remapping. Like for BBN, the arguments leading to this parameter combination are based on the assumption that the specific entropy remains constant. The CRR produces a total of $\simeq 5.4$ photons per hydrogen nucleus, implying that the recombination process indeed changes the specific entropy of the Universe \citep{Chluba2006FF, Chluba2010He}. However, since the baryon number is extremely small (compared to the photons), this correction can be neglected.

\vspace{-3mm}
\section{Effects of inhomogeneous recombination}
\label{sec: var}
%------------------
The discussion presented in the previous sections was based on the assumption that the recombination process proceeds in the same way everywhere. However, even within \LCDM, small fluctuations in the CRR are expected due to the presence of cosmological perturbations. Because of the non-linear dependence of the recombination process in particular on the local CMB monopole temperature \citep[see e.g.,][]{Chluba2008There}, this can lead to variations that modify the CRR at a small level once averaged over the sky.

Although a rigorous computation of this effect is beyond the scope of this paper, in the separate Universe approximation we can estimate the effect by considering the modifications to the CRR when varying the local cosmological parameters. Assuming that all the recombination radiation arises from their respective recombination eras, we can, for example, estimate the effect of perturbations on the hydrogen recombination spectrum as
%------------------------------------
\begin{align}
\label{eq: taylor 2}
\langle \Delta I^{\rm HI}_{\rm CRR} (p)\rangle 
&\approx \Delta I^{\rm HI}_{\rm CRR}(\bar{p}) 
+ \frac{1}{2}\sum_{i,j}\frac{\partial^2 (\Delta I^{\rm HI}_{\rm CRR})}{\partial \ln \bar{p}_i\partial \ln \bar{p}_i}
\,\left< \frac{\Delta p_i}{\bar{p}_i} \frac{\Delta p_j}{\bar{p}_j}\right>.
\end{align}
%------------------------------------
Here, $\langle \ldots \rangle$ denotes the cosmological average across the sky and $\bar{p}$ determines the average parameters. Importantly, while the linear order derivative term drops out after the ensemble average is carried out, at second order in the parameter variations a modification of the CRR is expected. 

The HI radiation originates from $z\simeq 1400$ \citep{Sunyaev2009Signals}, which implies that the HI contribution to the CRR is sensitive to the amplitude of fluctuations at this redshift. Similarly, the HeI and HeII distortion arise from $z\simeq 2000$ and $z\simeq 6000$, probing the amplitude of fluctuations at even earlier times. Evidently, by the time we observe the CRR, at sub-horizon scales these earlier fluctuations have been altered (e.g., by have Silk damping), but the effect on the average CRR will still be present almost unaltered since thermalization processes are already inefficient \citep{Burigana1991Formation, Hu1993ThermalizationI}. This implies that imprints of the invisible, pre-recombination perturbations are still visible in the average CRR spectrum, while at super-horizon scales the fluctuations may still be visible directly as CRR anisotropies due to patch to patch variations.

%------------------------------------
\begin{figure}
    \centering
    \includegraphics[width=\columnwidth]{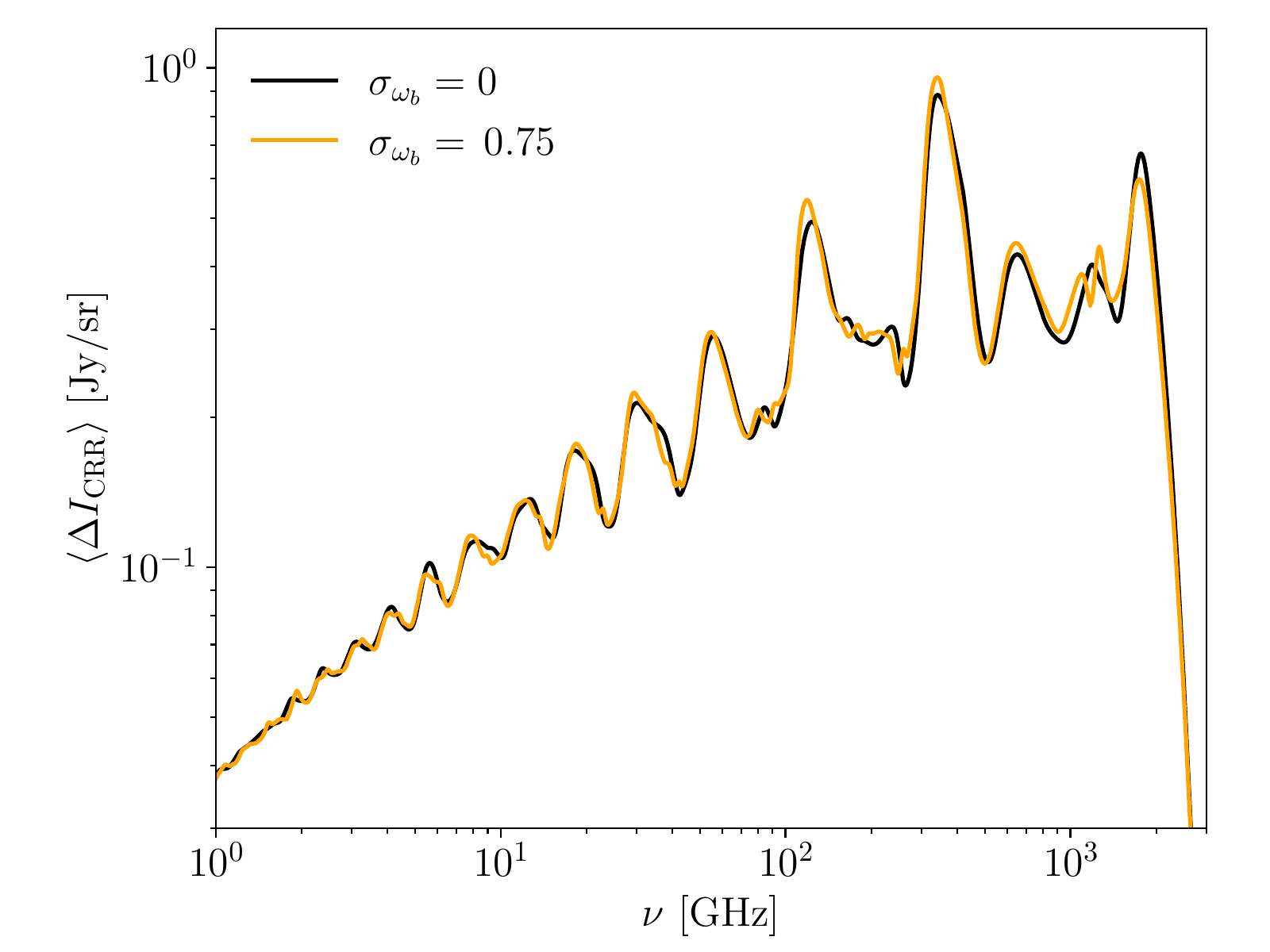}
    \caption{Sky-averaged CRR spectrum with (orange) and without (black) fluctuations of the baryon energy density for $\sigma_{\omb} \equiv \langle \Delta\omb^2 \rangle^{1/2}/\baromb= 0.75$. The variations in the baryon density lead to small shifts in the position of the CRR features as well as broadening of the lines.}
    \label{fig: inhom_from_Taylor_b}
\end{figure}
%------------------------------------

%------------------------------------
\begin{figure}
    \centering
    \includegraphics[width=\columnwidth]{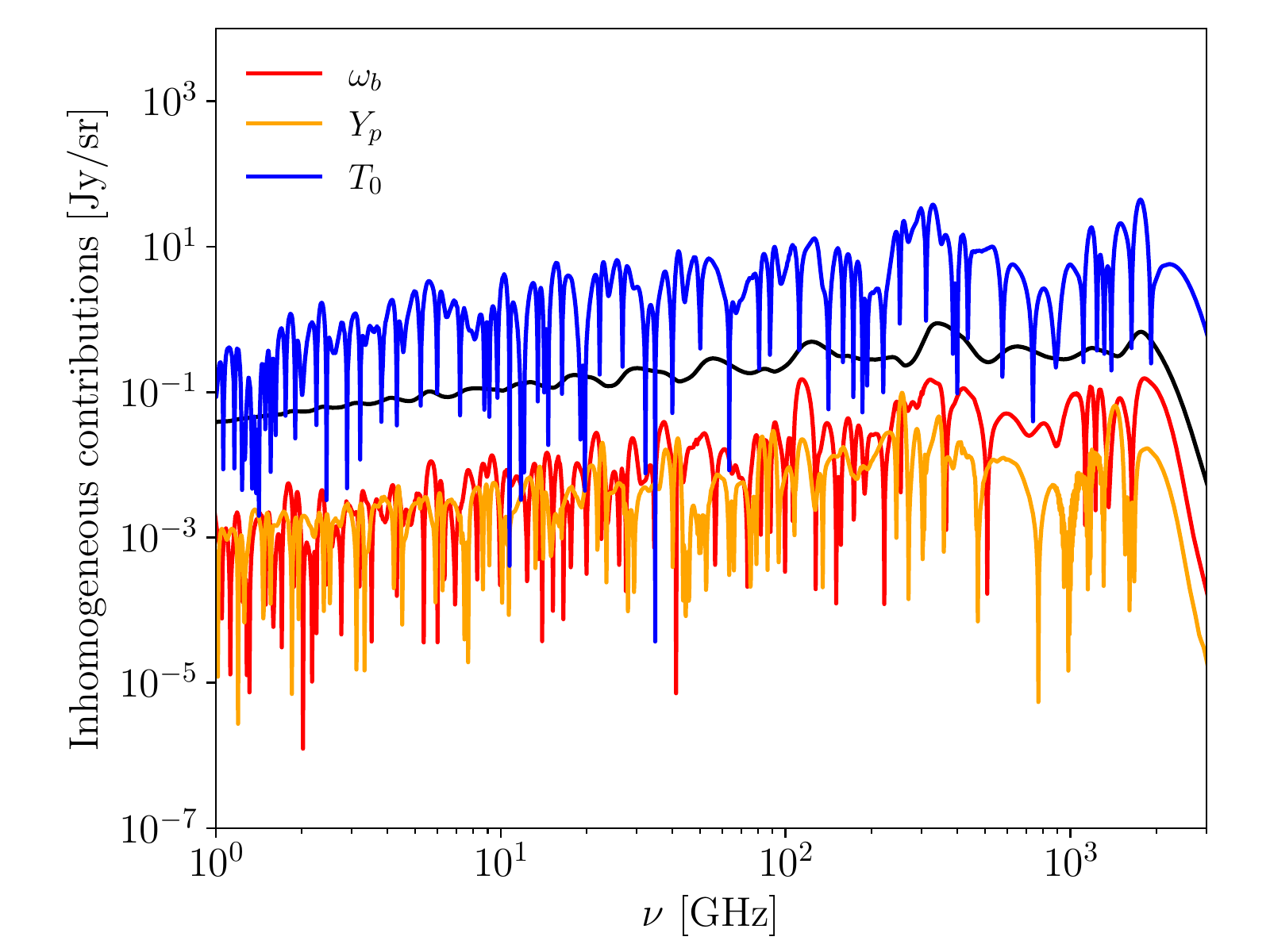} \\
    \includegraphics[width=\columnwidth]{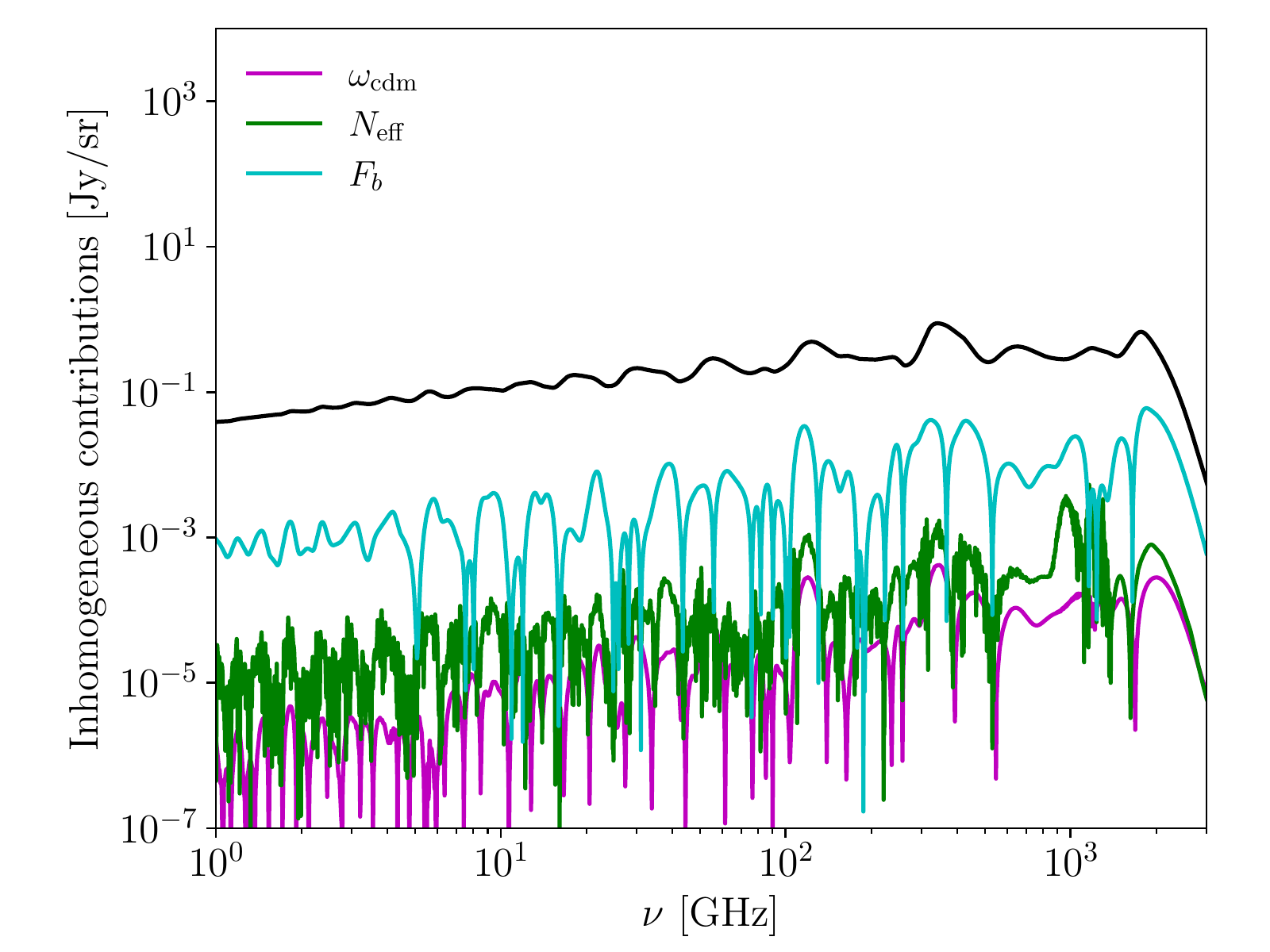}
    \caption{Illustration of the second-order spectral contributions for all considered parameters. The curves are equivalent to the parameter response spectra, $(1/2) \partial^2 \Delta I_{\rm CRR}/\partial \ln \bar{p}^2$. For reference we show the average CRR in black. Fluctuations in the cosmological parameters lead to a rich spectral response structure that in principle should allow us to separate the contributions from various terms.}
    \label{fig: inhom_from_Taylor_full}
\end{figure}
%------------------------------------

In Fig.~\ref{fig: inhom_from_Taylor_b} we illustrate the effect of fluctuations in the baryon density on the CRR for $\sigma_{\omb} \equiv \langle \Delta\omb^2 \rangle^{1/2}/\baromb= 0.75$. The baryon density fluctuations introduce only modifications to the CRR at the level of $\simeq 1-10\%$, which is to be expected given the close to linear dependence of the CRR spectrum on $\omb$ (see Fig.~\ref{fig: DI_vs_params}). In Fig.~\ref{fig: inhom_from_Taylor_full} we illustrate the variance contributions to the CRR for all the other standard recombination parameters. In addition, we show how local baryon density enhancements $n_{\rm b}=F_{\rm b}\,n_{\rm b}^{\Lambda \rm CDM}$, relating to the effects of PMFs, could modify the CRR at second order (see Sec.~\ref{sec: Hubble} for more details).

Figure~\ref{fig: inhom_from_Taylor_full} clearly indicates that inhomogeneities in $T_0$ have by far the largest effect. This is directly followed by the responses with respect to $\omb$ and $F_{\rm b}$, while all other parameters have a suppressed effect. For variations in $T_0$, this is not too surprising, given the exponential dependence of the recombination process on this parameter. For $\omb$ (and similarly also for $F_{\rm b}$) variations, although the first-order coefficients are much larger than the second-order ones (see Fig.~\ref{fig: DI_vs_params}), the dependence of the CRR on $\omb$ is large enough for relatively small second-order corrections to stay relevant. 
The response to $Y_p$ is suppressed by the fact that at second order only the subdominant HeI and HeII contributions are affected.
For the remaining parameters, the effect is suppressed since these parameters mainly enter indirectly through the expansion rate.

Since deviations of the CRR from the reference of the order of a few percent are expected to be measurable with experiments such as Voyage 2050+ \citep{Hart2020Sensitivity}, these effects may indeed become observable. This is particularly interesting because, although large levels of variation are not expected within \LCDM (see Sec. \ref{sec: LCDM}), they might be produced in non-standard models (e.g., see Sec. \ref{sec: Hubble}). In the next sections, we will consider a number of cases highlighting what one may be able to learn using future CMB spectroscopy.

%------------
\begin{figure*}
    \centering
    \includegraphics[width=0.47\textwidth]{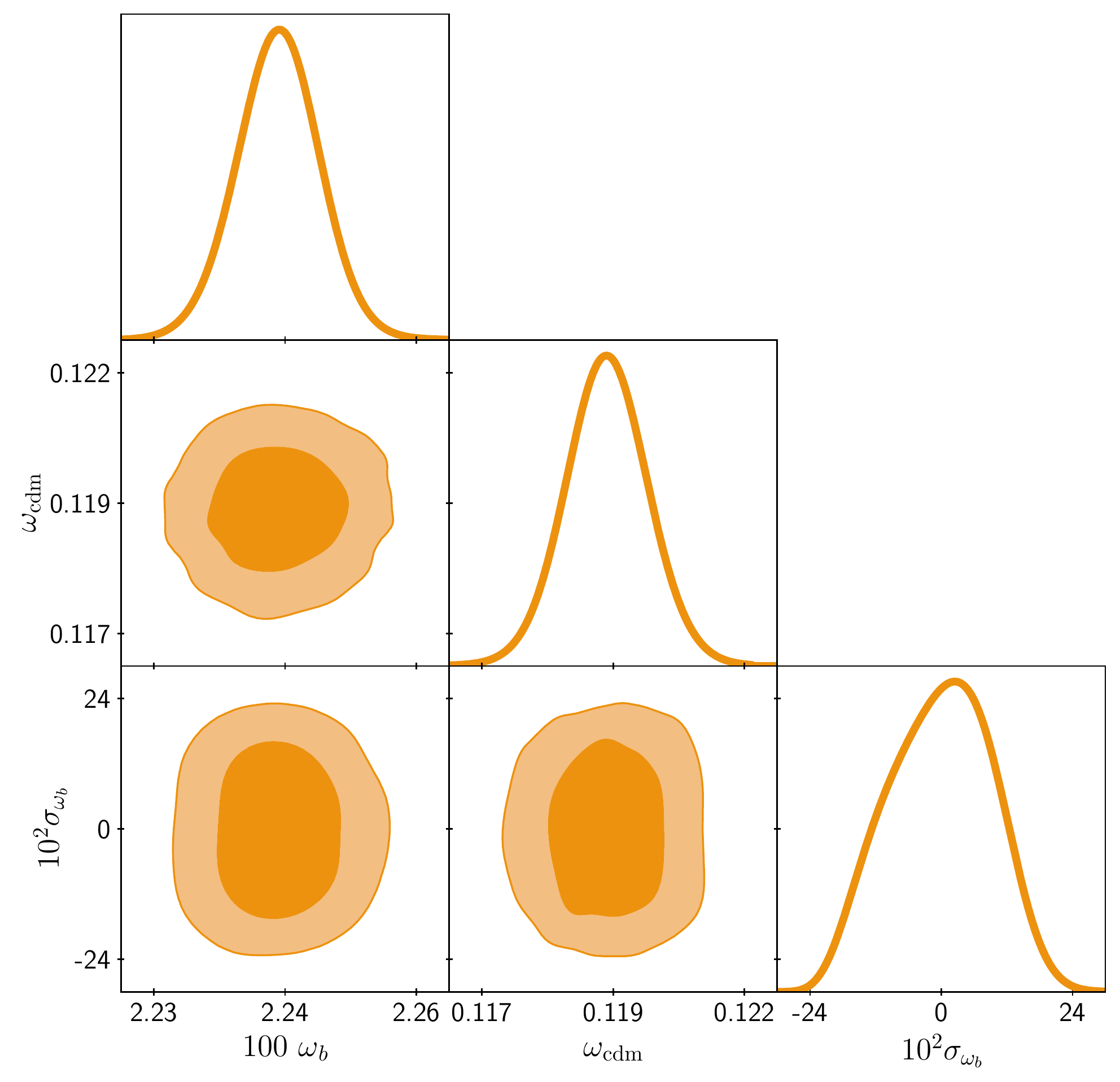}
    \includegraphics[width=0.47\textwidth]{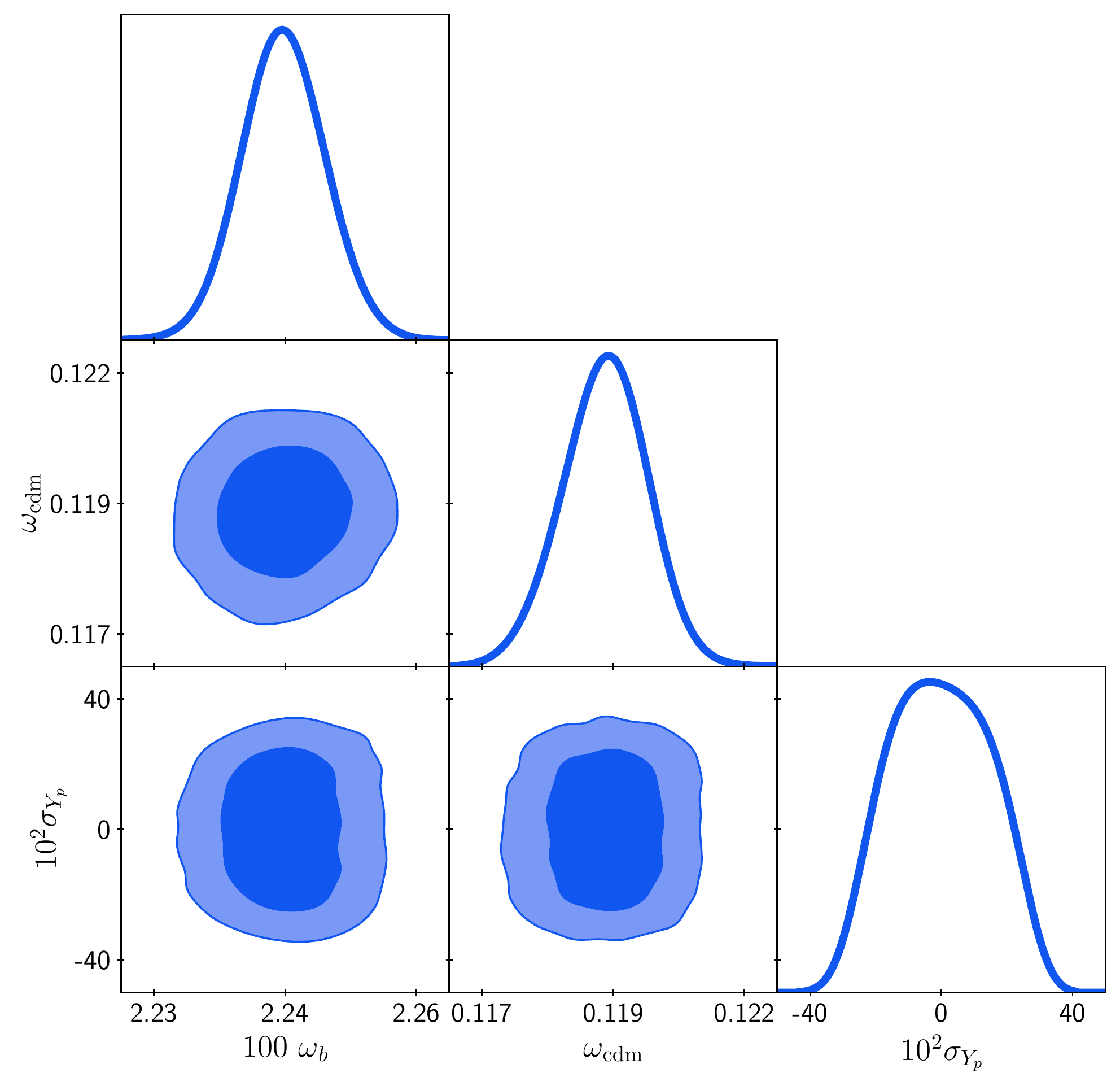}
    \includegraphics[width=0.47\textwidth]{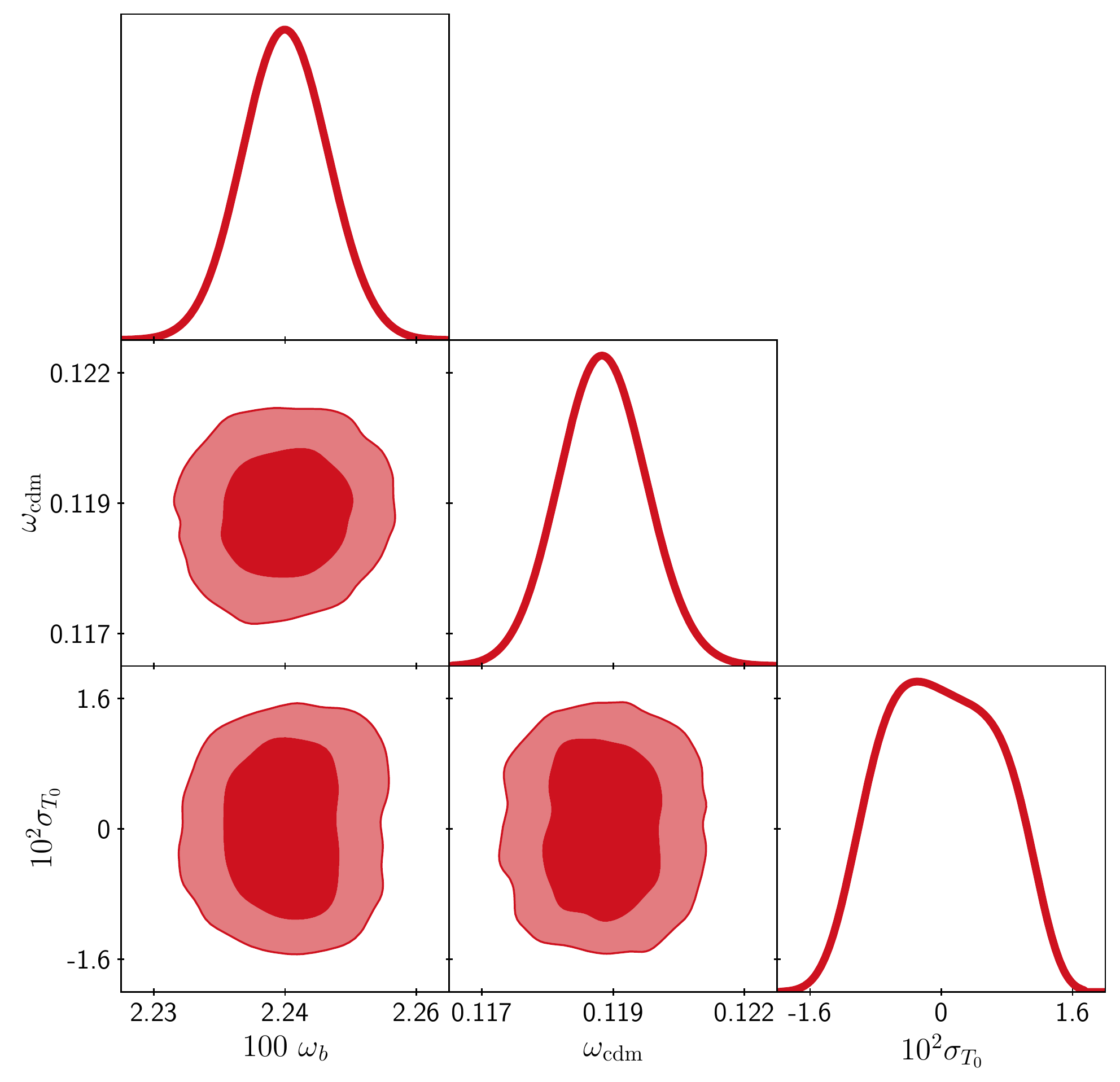}
    \includegraphics[width=0.47\textwidth]{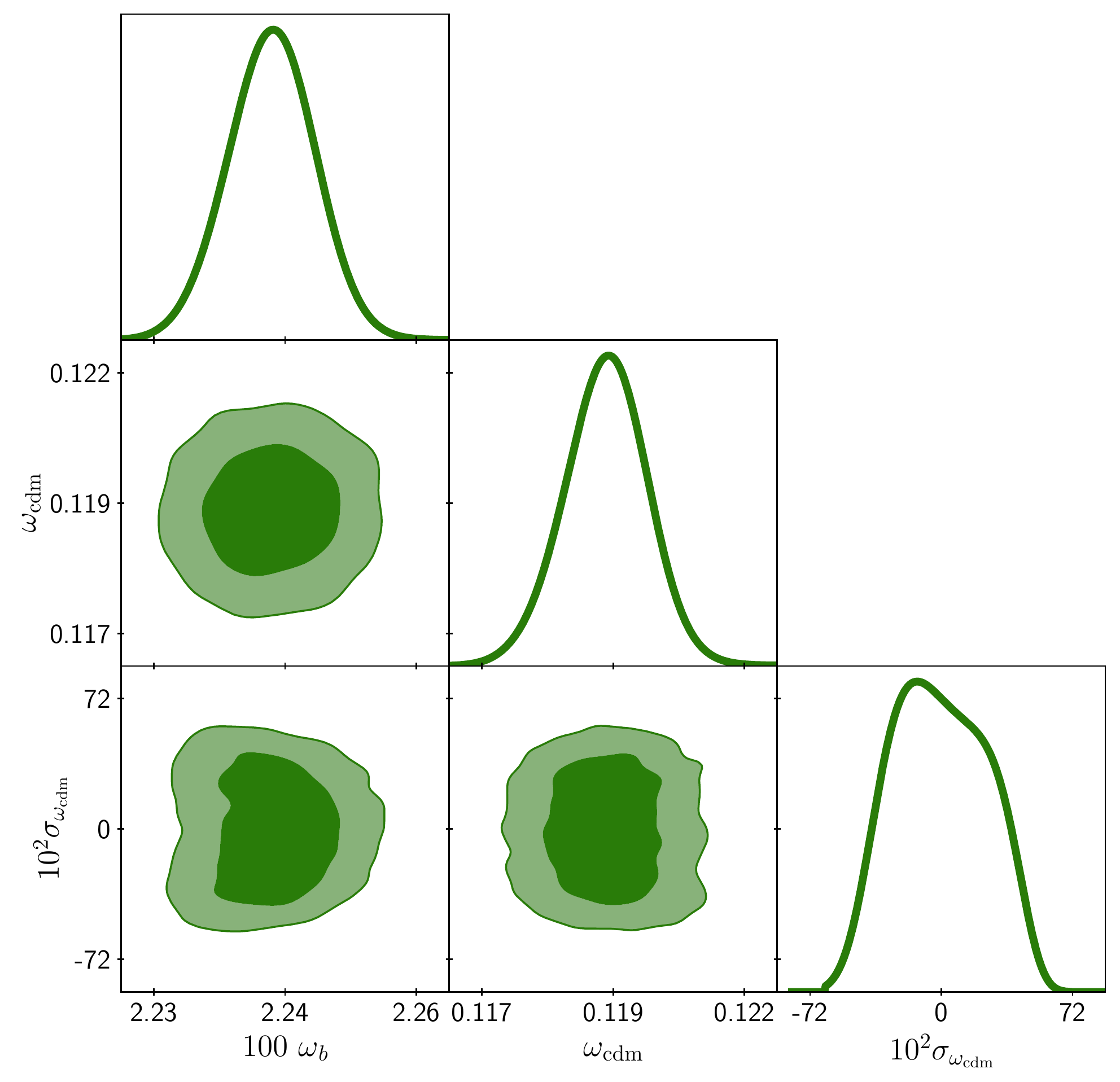}
    \caption{1D posterior and 2D contours (at 68\% and 95\% CL) of the variances of $\omb$, $Y_p$, $T_0$ and $\omega_{\rm cdm}$ (from top to bottom, left to right) assuming a Voyage 2050+ mission.}
    \label{fig: MCMC_res_1p}
\end{figure*}
%------------

\subsection{Detectability forecast}\label{sec: detect}
%------------
As a first point, we investigate to which degree parameter variance contributions to the CRR could be measured with future SD missions. To quantitatively assess the situation, we perform parameter scans similar to those performed in Sec.~\ref{sec: fore} including parameter variations $\sigma_{i,j}^2=\langle (\Delta p_i/\bar{p}_i) (\Delta p_j/\bar{p}_j) \rangle$, as given in Eq.~\eqref{eq: taylor 2}. 
We assume that the $\sigma_{i,j}$ are independent of redshift, even if more generally one could introduce independent parameters for the three recombination eras.
Since these parameters only affect the CRR contribution to the total SD spectrum, for sake of computational simplicity we neglect the contribution from other sources of SDs and include \Planck priors on all \LCDM parameters as well as on $Y_p$ and $N_{\rm eff}$ when necessary (these parameters are otherwise fixed to the reference values given in Sec.~\ref{sec: cosmo_dep}, and so is $T_0$). We also do not marginalize over any foreground, leaving a more comprehensive optimization to the future. For sake of brevity, we focus on the Voyage 2050+ setup.

We start our discussion with single parameter extensions including the variance of  $\omb$, $Y_p$, $T_0$ and $\omega_{\rm cdm}$ only. In Fig.~\ref{fig: MCMC_res_1p} the posteriors of the corresponding runs are presented. For $T_0$, the spectrometer itself will provide an unprecedented measurement and even though in scenarios with inhomogeneous BBN one would also anticipate that the average value of $Y_p$ would have to be independently constrained, we made this choice for illustration.
We can see that measurements of the CRR can be quite sensitive to variance contributions. Quoting $\sigma_p \equiv \langle \Delta p^2 \rangle^{1/2}/\bar{p}$ for the individual parameters, we find sensitivities at the 1$\sigma$ level of $11\%$, $17\%$, 0.7\% and $29\%$ for $\omb$, $Y_p$, $T_0$ and $\omega_{\rm cdm}$, respectively. As we check explicitly, varying all variances at the same time does not significantly reduce the constraining power of the CRR with respect to these quantities. To a very good approximation, we find that cross-variances, such as e.g., $\sigma_{\omb\omega_{\rm cdm}}$, can be approximated by the square-root of the product of the single variances, i.e., $\sigma_{\omb\omega_{\rm cdm}}\simeq(\sigma_{\omb}\sigma_{\omega_{\rm cdm}})^{1/2}\simeq 18\%$.\footnote{This is because the CRR response to the underlying parameters is very different (see Fig.~\ref{fig: DI_vs_params}), with little cross-correlation between the derivatives.}
We note that if indeed large variance contributions are observed, it may be necessary to also look for higher order corrections to the CRR. These can in principle be added using higher order Taylor terms, which give further insights into the higher-order statistics of the fields.

\subsection{\LCDM contribution}
\label{sec: LCDM}
%------------------------------------

\subsubsection{Primordial density fluctuations}
%------------------------------------
As mentioned above, already within $\Lambda$CDM the propagation of primordial density fluctuations $\delta(z,k)$ perturbs the otherwise homogeneous background and leaves an imprint on the CRR. Since the presence of such perturbations is intrinsic to the $\Lambda$CDM model, this sets an unavoidable minimum contribution to the sky-averaged CRR spectrum below which second-order contributions need to be taken into account for the modeling of the CRR. The contributions to the average spectrum should persist as witnessed during the respective recombination era, even if the related perturbations have long been modified.

To compute the $\Lambda$CDM contributions [i.e., the second order term in Eq.~\eqref{eq: taylor 2}], one would need to follow the time evolution and scale dependence of the density perturbations for all fluids relevant around recombination (i.e., DM, baryons, photons and neutrinos) between the first helium recombination (when the CRR begins) and the end of hydrogen recombination (when the CRR ends), thereby spanning the approximate redshift range $z\simeq 500-8000$. 
At every redshift it then becomes possible to calculate the ensemble average over the sky of the perturbations as\begin{align}\label{eq: delta_delta}
\langle \delta_i \delta_j \rangle (z) = \int \mathcal{P}(k) \delta_i (z,k) \delta_j (z,k) \text{d}\ln k\,,
\end{align}
where $\mathcal{P}(k)$ is the dimensionless PPS. 
Here, $\delta_i=\Delta p_i/\bar{p}_i$ for each of the parameters (assuming linearizable perturbations).
Integrating over redshift the various combinations $\langle \delta_i \delta_j \rangle$ allows us to obtain the total contribution for each individual stage of the recombination process (i.e., HI, HeI, HeII) to be added to the unperturbed CRR spectra. The final result is the sum of all contributions.

Since the characteristics of these perturbations depend on the unknown initial conditions of the universe, the shape of this \LCDM component will inevitably inherit some model dependence. Furthermore, although the shape of the PPS is known at scales below $k\simeq 1$~Mpc$^{-1}$ \citep{Akrami2018PlanckX}, this is not the case at smaller scales where only upper bounds exist (see e.g., Fig. 8 of \cite{Byrnes2018Steepest} for a graphical overview). This means that eventual deviations of the PPS from the \Planck estimates would introduce another layer of uncertainty in the determination of these terms.

Nevertheless, it is very instructive to consider some benchmarking cases to illustrate the overall magnitude of this contribution. As a proof of principle, here we will solely focus on adiabatic initial conditions and on the redshift at which the different contributions to the CRR originate, i.e., $z=6000$ (HeII), $z=2000$ (HeI) and $z=1400$ (H), without performing any involved redshift-dependent integration, which we leave for future work. The integral of Eq.~\eqref{eq: delta_delta} is performed between scales $k=10^{-4}-1$~Mpc$^{-1}$, above which non-linear effects become important. The resulting values of the $\langle \delta_i \delta_j \rangle$ combinations are all $\simeq 5\times 10^{-8}$ at all redshifts, except for $\langle \delta^2_{\rm cdm}\rangle$ which is approximately $3\times10^{-6}$ during helium recombination and $1\times10^{-5}$ during hydrogen recombination. Nevertheless, since the typical contribution from DM fluctuations to the CRR is roughly two orders of magnitude lower than that of the baryons (see Fig. \ref{fig: inhom_from_Taylor_full}), the overall impact of the DM variations on the CV term does not exceed that of the baryons. 

In terms of $\sigma_p\simeq\langle \delta_p^2 \rangle^{1/2}$, the most relevant contributions (as compared to the sensitivities mentioned in the previous section) are then given by $\omega_{b}$, $T_0$ and $\omega_{\rm cdm}$ which lead to $\sigma_{\omega_{b}}\simeq 0.02\%$, $\sigma_{T_0}\simeq 0.006\%$ (using $\Delta T/T\simeq \delta_\gamma/4$) and $\sigma_{\omega_{\rm cdm}}\lesssim 0.3\%$, respectively. Since these lay two orders of magnitude (or more) below the detectability threshold of Voyage 2050+, the impact that the primordial spacial fluctuations of \LCDM have on the CRR can be safely neglected. Nevertheless, even for \LCDM, spatial variations in the CRR could open a new way to probe the growth of structures in extremely early phases, and thus may warrant further study, possibly by extending recently developed tools for computing CMB SD anisotropies \citep{SS2023II, SS2023III}.

\subsubsection{Small scale CMB temperature fluctuations}
\label{sec:T0var}
%------------------------------------
It is well-known that the mixing of blackbodies of different temperatures sources $y$-type distortions \citep[e.g.,][]{Chluba2004Superposition, Stebbins2007CMB}. In the early universe, this process is indeed at work due to the dissipation of small-scale acoustic perturbations \citep{Sunyaev1970SmallI, Daly1991Spectral, Hu1994Power, Chluba2012CMB}. At $z\lesssim 10^{4}$, all perturbations in the CMB temperature will contribute to an average $y$-type distortion with $y\simeq \frac{1}{2}\langle \delta_T^2 \rangle$. Usually, this contribution is significantly smaller than that caused by the localized heating in SZ clusters and the hot gas filling the universe at low redshifts, with an expected total $y\simeq 2\times 10^{-6}$ \citep{Hill2015Taking}. In terms of an effective CMB temperature variance, this implies $\sigma^*_{T}\simeq 0.2\%$. As we saw above, with a detailed measurement of the CRR we may be able to reach a sensitivity of $\sigma_{T}\simeq 0.7\%$, implying that we could confirm if there has been any contributions to the effective $y$-parameter from primordial small-scale fluctuations at a level that is larger than the late-time cluster contribution.

\subsection{PMFs and Hubble tension}
\label{sec: Hubble}
%------------------------------------
One possible avenue towards solving the persistent Hubble tension could be related to small-scale density perturbations that are introduced by PMFs \citep[e.g.,][]{Jedamzik2020Relieving, Galli:2021mxk}. If the small-scale baryon density perturbations are indeed present during the various recombination eras, then this inevitably leaves an imprint in the average CRR. This statement is true even if after the recombination process has completed the perturbations are completely erased by photon diffusion, since the spatial structure of the distortion signal is irrelevant for the monopole spectrum.

To illustrate the possible effects, we follow \citet{Jedamzik2020Relieving, Galli:2021mxk} by introducing a three recombination zone model with baryon density enhancements of $F^{(1)}_{\rm b}=0.1$ and  $F^{(2)}_{\rm b}=1$ for two of the three zones, and volume filling factor $f^{(2)}_V=1/3$ for the second zone. We then consider two models with baryon density variance $b\equiv \sigma^2_{\rm b}=0.5$ and $3$, which with 
%------------------------------------
\begin{align}
\label{eq:F_b}
\sum f_V^{(i)}=1,
\quad
\sum F^{(i)}_{\rm b} f_V^{(i)}=1,
\quad
\sum \left[F^{(i)}_{\rm b} \right]^2 f_V^{(i)}=1+b
\end{align}
%------------------------------------
determines the full solution. The local baryon density is given by $n^{(i)}_{\rm b}=n_{\rm b}^{\Lambda{\rm CDM}}\,F^{(i)}_{\rm b}$ and we assume that the Hubble expansion rate is not affected by the modification.
For the two cases, we then find  $F_{\rm b}=(0.1, 1, 6)$ with $f_V=(0.566, 0.333, 0.101)$ for $b=3$ and $F_{\rm b}=(0.1, 1, 1.833)$ with $f_V=(0.321, 0.333, 0.346)$ for $b=0.5$. These specific scenarios have been shown to play a relevant role in the context of the Hubble tension.

%------------------------------------
\begin{figure*}
    \centering
    \includegraphics[width=\columnwidth]{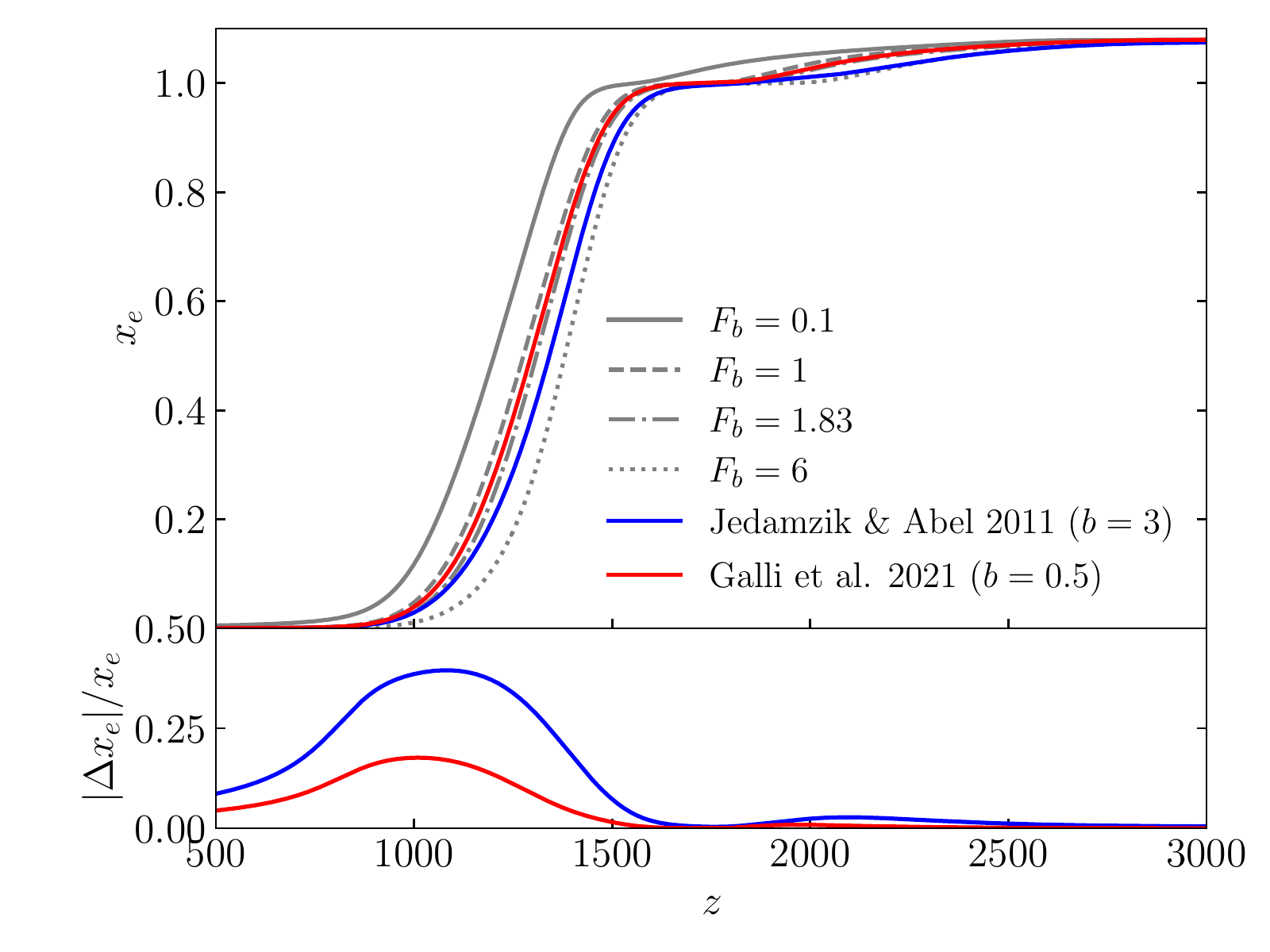}
    \includegraphics[width=\columnwidth]{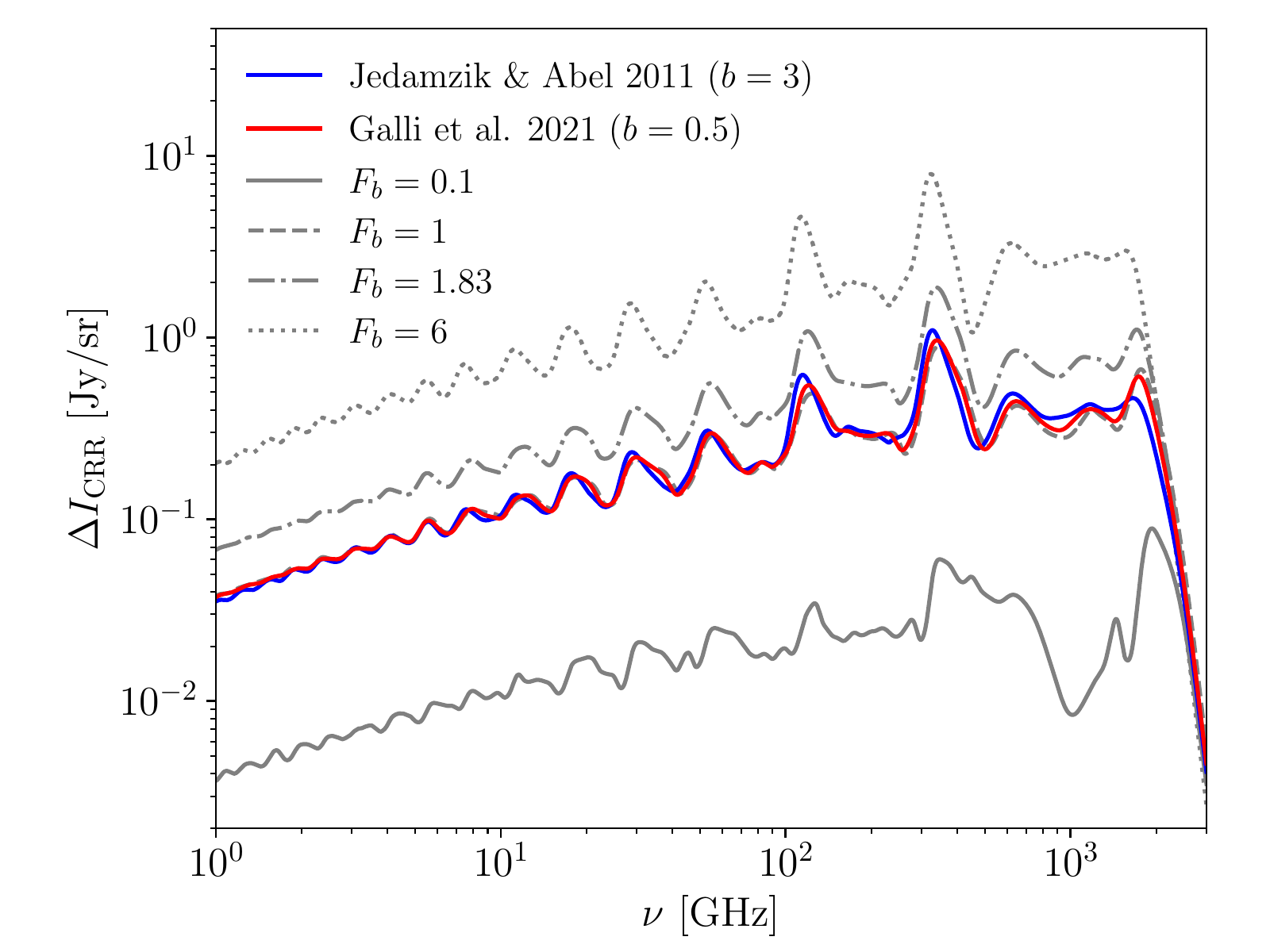}
    \caption{Recombination histories (left panel) and CRR spectra (right panel) for the PMF models considered in the text. The individual spectra for various values of the baryon density enhancement factor are shown as gray lines (the $F_{\rm b}=1$ case corresponds to \LCDM). The averages \citep[chosen following examples of][]{Jedamzik:2013gua, 
    Jedamzik2020Relieving, Galli:2021mxk} are instead displayed in blue and red, and show significant second order contributions, manifesting in smearing of the lines and shifts in their position.}
    \label{fig: PMFs}
\end{figure*}
%------------------------------------
In Fig.~\ref{fig: PMFs} we show the resulting electron recombination histories (left) and CRR spectra (right). The gray lines represent the individual spectra for the various aforementioned values of the baryon density enhancement factor, with the $F_{\rm b}=1$ case corresponding to \LCDM. The blue and red curves represent the average recombination histories following the results of \cite{Jedamzik2020Relieving} ($b=3$) and \cite{Galli:2021mxk} ($b=0.5$), respectively. 
These were computed explicitly using {\tt CosmoRec} \citep{Chluba2011Towards}, which accurately treats the hydrogen and helium recombination problem, including subtle transfer effects \citep{Chluba2012HeRec}.
One can clearly see that the second order variations lead to a non-trivial broadening and shifts in both cases. 
For the electron recombination process, underdense regions recombine later, while overdense regions recombine faster. In a similar manner, underdense regions have a smaller total emission in the CRR, with the line positions shifted towards higher frequencies (i.e., less redshifting after the later emission), and \textit{vice versa} for overdense regions. The modifications are {\it asymmetric} around the means, implying the non-trivial CRR modifications.

%------------------------------------
\begin{figure}
    \centering
    \includegraphics[width=0.48\textwidth]{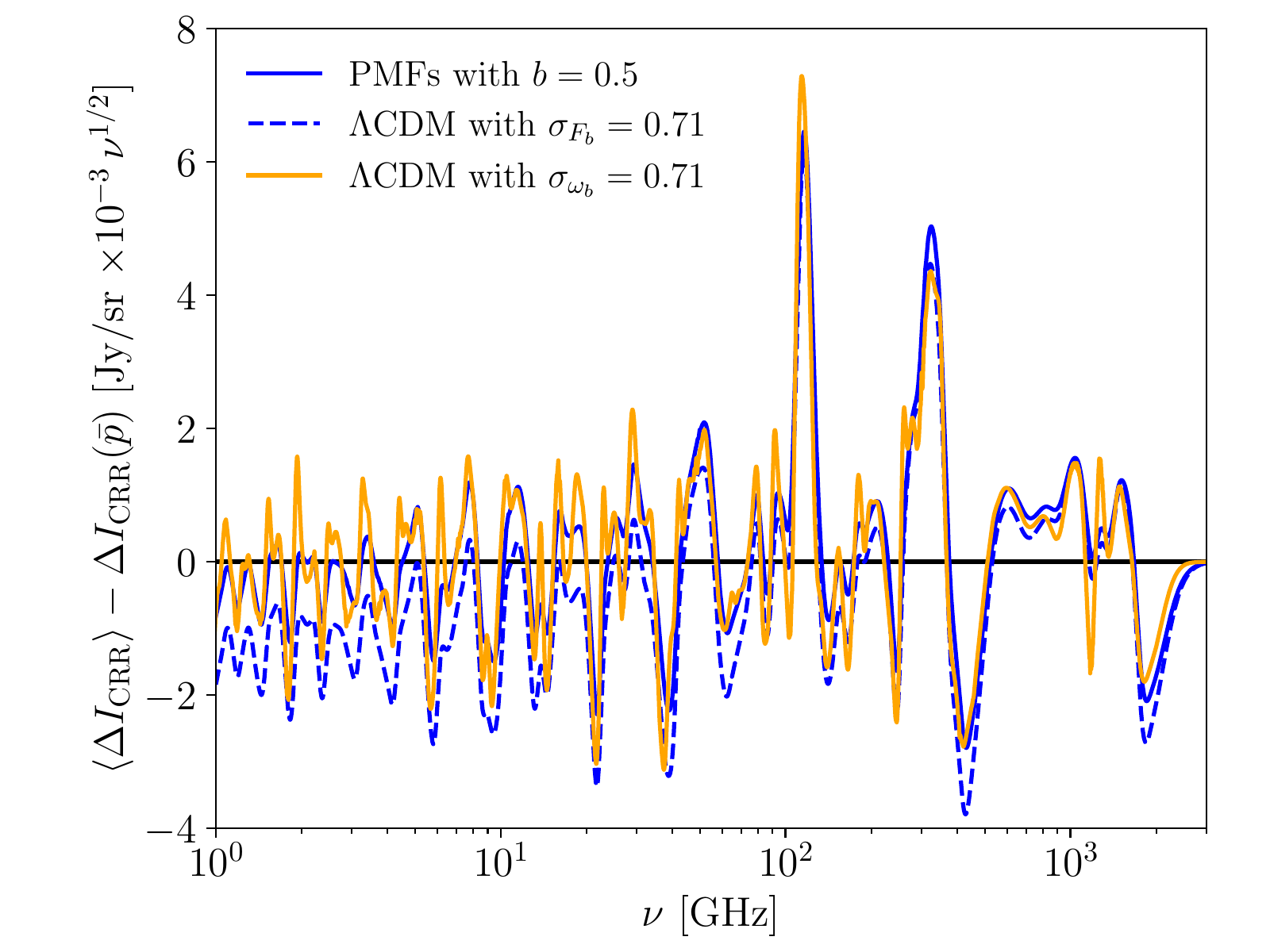}
    \\
    \caption{Illustration of the differences in the CRR from fluctuations in the baryon density and baryon density enhancements. The non-linear averaging process is consistent with the expectations and variance in $\omb$ mimics the effect of variance in $F_{\rm b}$.}
    \label{fig:PMFs_vs_sigmab}
\end{figure}
%------------------------------------
In Fig.~\ref{fig:PMFs_vs_sigmab} we illustrate that the modifications to the CRR spectrum (blue solid curve) can be approximately captured by considering the second order derivative of the CRR spectrum in terms of $F_{\rm b}$ (blue dashed curve). The direct difference is in fact well approximated by $\sigma^2_{F_{\rm b}}\equiv b$. Furthermore, for comparison we also show the impact of the variance of $\omb$ (orange solid curve). Physically, the role of the variances of $F_{\rm b}$ and $\omb$ has a slightly different meaning, since the expansion rate is also affected in the latter case. However, overall many of the spectral features are similar and one can expect the sensitivity to the variance of $F_{\rm b}$ to be similar to the variance of $\omb$. As a consequence, since Voyage 2050+ would be sensitive to $\sigma_{\omb}\simeq 10\%$, it would also be able to test clumping factors $b$ as low as 0.01, thereby fully exploring the region of parameter space relevant for the Hubble tension. This would open the way for confirming the physical origin of the Hubble tension should it be related to inhomogeneous recombination scenarios. 

As a final comment, even beyond the Hubble tension, this type of constraints would impose some of the most stringent limits on the PMF strength to date (estimated to be of the order of the~pG). A similar analysis could also be performed in the context of inhomogeneous BBN \citep{PhysRevD.64.023510, 2010arXiv1007.0466N} and all other models leading to baryon clumping at small scales. 
In this respect, constraints on variation of $Y_p$ derived from the CRR could supersede those currently possible from stellar abundance measurements \citep[e.g., see][]{Arbey2020}. Most importantly, the constraints on $Y_p$ from the CRR should not be prone to any late time stellar physics effects, providing the most primordial constraint one could hope for.
A dedicated analysis is however left for future work.

\vspace{-3mm}
\section{Conclusions}\label{sec: conc}
%------------------------------------

The CRR is a SD of the CMB energy spectrum sourced by the absorption/emission of photons during the cosmological recombination process. Because of that, its shape is very featureful and remains unaffected by the thermal history post-recombination. Therefore, the CRR is the most direct probe of the recombination process we could have access to. As such, it can be used to tightly constrain deviations from the standard picture, which have become particularly interesting in light of the Hubble tension.

Because of this potential role that the CRR could play in modern cosmology, the development of appropriate numerical tools becomes necessary. For this reason, based on the accurate \texttt{CosmoSpec} code, the main contribution of this work to the topic is the development of a new emulator, named \texttt{CRRfast}, designed for the fast and accurate computation of the CRR spectrum. 

Concretely, as extensively explained in Sec.~\ref{sec: num}, \texttt{CRRfast} relies on a second-order Taylor expansion of the CRR spectrum around a fiducial model (arbitrarily set to \LCDM with \Planck best-fitting values) for all relevant cosmological parameters (overviewed in Sec.~\ref{sec: cosmo_dep}). The resulting Taylor coefficients (representing the response of the spectrum to a given variation of a parameter) can then be tabulated and used to calculate the CRR spectrum for any set of values of the aforementioned quantities. The reference spectrum and its parameter-dependent variations are computed with \texttt{CosmoSpec}, which \texttt{CRRfast} effectively emulates.

The main advantage of \texttt{CRRfast} with respect to \texttt{CosmoSpec} is a more than 500-fold reduction of the computation time, bringing one CRR evaluation to sub-seconds. Furthermore, although so far \texttt{CRRfast} only covers the \LCDM model and its minimal extensions, it can be easily expanded to include also other cosmological models (many of which are already implemented in \texttt{CosmoSpec} and whose addition would be straightforward). The only requirement is that the impact of the underlying parameters on the CRR can be captured by the Taylor expansion. 
To extend the valid parameter region, multiple Taylor pivots can be used.
We note that \texttt{CRRfast} has been made publicly available both as a stand-alone code and as part of the Boltzmann solver \texttt{CLASS}. The latter is particularly relevant since the inclusion of the CRR in the pipeline completes the set of \LCDM sources of SDs accounted for by the code. 

In this way, \texttt{CRRfast} is ideal for statistical analyses and opens the way for the systematic cosmological exploration of the CRR, while \texttt{CosmoSpec} remains fundamental for precision calculations and the inclusion of new physics as well as new cosmological models. Many of such possible applications of \texttt{CRRfast} are highlighted in Sec.~\ref{sec: app}. For instance, by means of the newly-developed \texttt{CLASS} implementation, here we perform the first all-inclusive (in terms of considered effects) forecast for the constraining power of CMB SDs within \LCDM. This illustrates that CMB SDs could independently constrain four of the six \LCDM parameters, namely $\omb$, $\omega_{\rm cdm}$, $n_s$ and $A_s$, at a statistically significant level with a Voyage 2050 mission. This wide-ranging constraining power is second only to the CMB anisotropy power spectra and sets the stage for further studies in this direction. In combination with CMB anisotropy data this opens the path for improved constraints but also for novel tests of the \LCDM model, an analysis that is now possible and planned for future work.

Another interesting avenue opened by the development of \texttt{CRRfast} involves design studies for missions targeting the CRR (and CMB SDs more in general), following up on \citet{SathyanarayanaRao:2015vgh, Vince2015Detecting, Abitbol2017Prospects, Chluba2019Voyage}. As a simple example, here we explicitly focus on Voyage 2050 and in particular on its LFM, finding that for the CRR it might not deliver a significant improvement over a setup without it. Since the LFM poses one of the significant cost-drivers and technological challenges, a future systematic cost-reward analysis in this direction would be of major interest. However, it is also clear that a reduced sensitivity at low frequencies could hamper the performance with respect to primordial $\mu$-distortions \citep{Abitbol2017Prospects}, thus requiring a more careful optimization that also includes the effects of spatial variations of foregrounds \citep[e.g.,][]{Rotti2021} in a more complete manner.

In Sec.~\ref{sec: var} we illustrate how {\tt CRRfast} naturally opens the path for investigations of spacial variations of the relevant cosmological parameters during the recombination era. Due to non-linear effects, these broaden and shift the spectral features of the CRR. Because of the primordial fluctuations existing within \LCDM, the presence of these second-order contributions is guaranteed and their observability can now be assessed with \texttt{CRRfast}. Of course, also many beyond-\LCDM scenarios predict such variations, as in the presence of inhomogeneuos BBN and PMFs (which generate clumps of matter at small scales), possibly enhancing the inhomogeneous contributions in a significant way. Since PMFs have been proposed as a possible solution to the Hubble tension, this further highlights the role that the eventual observation of CMB SDs could play in the context of the Hubble tension.  

Here we focus on the representative cases of \LCDM and PMFs. We find that, while the \LCDM contribution would be out of reach even for Voyage 2050+ sensitivities, such a mission would be able to fully explore the region of parameter space relevant for PMFs as a solution to the Hubble tension. Quantitatively, we forecast a Voyage 2050+ mission to be able to probe clumping factors of the order of $b\simeq0.01$, corresponding to PMF strengths of the order of pG. These would represent some of the most stringent limits on PMFs and their evolution to date. A more refined analysis studying the synergy between these bounds and those inferred from other cosmological effects of PMFs on observables such as the CMB anisotropy power spectra is left for future work.

In summary, \texttt{CRRfast} is a fast, easily extendable and publicly-available code for the computation of the CRR spectrum. In particular its \texttt{CLASS} implementation opens the door for the exploration of many interesting perspectives to be looked forward to in the future. This work highlights many of them and sets the stage for these up-coming analyses.

\section*{Acknowledgements}
We thank Karsten Jedamzik for the useful inputs on the manuscript.
ML is supported by an Fonds de la Recherche Scientifique de Belgique (F.R.S.- FNRS) fellowship.
This work was supported by the ERC Consolidator Grant {\it CMBSPEC} (No.~725456).
JC was furthermore supported by the Royal Society as a Royal Society University Research Fellow at the University of Manchester, UK (No.~URF/R/191023).
Computational resources have been provided by C\'ECI, funded by the F.R.S.- FNRS under Grant No. 2.5020.11 and by the Walloon Region.

\appendix
\section{Supplementary material on the Taylor expansion approximation}\label{app: details_Taylor}
%------------------------------------
The first derivative of Eq. \eqref{eq: taylor} can be simply expressed as
%------------------------------------
\begin{align}
    \frac{\partial(\Delta I_{\rm CRR})}{\partial p_{i}}\Bigg|_{\rm ref} \approx \frac{\Delta I_{\rm CRR}(p_{i, \rm ref}+\Delta p_{i})-\Delta I_{\rm CRR}(p_{i, \rm ref}-\Delta p_{i})}{2\Delta p_{i}} \,,
\end{align}
%------------------------------------
where $\Delta p_{i}=|p_i-p_{i,\rm ref}|$. In a similar way, one can define the second derivative for parameter combination $p_i p_j$ as
%------------------------------------
\begin{align}
    \nonumber \frac{\partial^2(\Delta I_{\rm CRR})}{\partial p_{i} \partial p_j}\Bigg|_{\rm ref} \approx & \frac{1}{4\Delta p_{i}\Delta p_{j}}[
    \Delta I_{\rm CRR}(p_{i, \rm ref}+\Delta p_{i},\, p_{j, \rm ref}+\Delta p_{j}) \\ 
    \nonumber & 
    -\Delta I_{\rm CRR}(p_{i, \rm ref}+\Delta p_{i},\, p_{j, \rm ref}-\Delta p_{j}) \\ 
    \nonumber & 
    -\Delta I_{\rm CRR}(p_{i, \rm ref}-\Delta p_{i},\, p_{j, \rm ref}+\Delta p_{j}) \\ 
    & +\Delta I_{\rm CRR}(p_{i, \rm ref}-\Delta p_{i},\, p_{j, \rm ref}-\Delta p_{j})]\,.
\end{align}
%------------------------------------
For the various $\Delta p_i$ we use $\Delta p_i/p_{i, \rm ref}=5\%$ for all parameters, noting that the resulting values for the derivative are largely insensitive to this choice. For the case of $T_0$, we find 1\% to deliver sufficiently accurate results for both small and large variations of this quantity.

\section{Supplementary material on the experimental setups}
\label{app: details_exp}
The considered experimental configurations are planned to be built out of three types of frequency modules, referred to as \text{low-,} mid-, and high-frequency modules (LFM, MFM and HFM, respectively). The LFM spans between 10 and 40 GHz with a 2.5 GHz bin width and represents the most sensitive of the three modules with a spectral sensitivity of \text{$2.9\times10^{-23}$ W/(m$^2$ Hz sr)} (assuming a 1-second integration). The MFM ranges between 20 and 600 GHz with a 20 GHz channel width and a sensitivity of \text{$1.2\times10^{-22}$ W/(m$^2$ Hz sr)} (assuming the same integration interval). Finally, the HFM extends the frequency array in the range between 400 and 6000~GHz with a 60 GHz channel width and a sensitivity of \text{$6.5\times10^{-23}$ W/(m$^2$ Hz sr)} (again with a 1-second integration).

According to \cite{Kogut2019CMB}, a \tquote{full} mission will be composed of four LFM, four MFM and one HFM, thereby reducing the aforementioned sensitivities by the same factors. We will furthermore assume the mission to collect data for a 8-years period, following~\cite{Chluba2019Voyage}, with the deployed configuration active for 25\% of the observational time, which was already the default for the original PIXIE configuration \citep{Kogut2011Primordial}. This reduces all sensitivities by an additional factor $\sqrt{8\,\text{yr}\times 0.25/\text{s}}\simeq 8\times10^3$, as clear e.g., from Eq.~(4) of \cite{Kogut2019CMB}. Explicitly, we have for the Voyage 2050 mission that the final sensitivities of the LFM, MFM and HFM are $1.8\times10^{-28},\, 7.6\times10^{-28}$ and $1.6\times10^{-27}$ W/(m$^2$ Hz sr), respectively. The corresponding values for Voyage 2050+ are simply rescaled by one order of magnitude.

\bibliographystyle{mnras}
\bibliography{bibliography}

\bsp
\label{lastpage}

\end{document}